\catcode`\@=11

\font\tenmsa=msam10
\font\sevenmsa=msam7
\font\fivemsa=msam5
\font\tenmsb=msbm10
\font\sevenmsb=msbm7
\font\fivemsb=msbm5
\newfam\msafam
\newfam\msbfam
\textfont\msafam=\tenmsa  \scriptfont\msafam=\sevenmsa
  \scriptscriptfont\msafam=\fivemsa
\textfont\msbfam=\tenmsb  \scriptfont\msbfam=\sevenmsb
  \scriptscriptfont\msbfam=\fivemsb

\def\hexnumber@#1{\ifnum#1<10 \number#1\else
 \ifnum#1=10 A\else\ifnum#1=11 B\else\ifnum#1=12 C\else
 \ifnum#1=13 D\else\ifnum#1=14 E\else\ifnum#1=15 F\fi\fi\fi\fi\fi\fi\fi}

\def\msa@{\hexnumber@\msafam}
\def\msb@{\hexnumber@\msbfam}
\mathchardef\boxdot="2\msa@00
\mathchardef\boxplus="2\msa@01
\mathchardef\boxtimes="2\msa@02
\mathchardef\square="0\msa@03
\mathchardef\blacksquare="0\msa@04
\mathchardef\centerdot="2\msa@05
\mathchardef\lozenge="0\msa@06
\mathchardef\blacklozenge="0\msa@07
\mathchardef\circlearrowright="3\msa@08
\mathchardef\circlearrowleft="3\msa@09
\mathchardef\rightleftharpoons="3\msa@0A
\mathchardef\leftrightharpoons="3\msa@0B
\mathchardef\boxminus="2\msa@0C
\mathchardef\Vdash="3\msa@0D
\mathchardef\Vvdash="3\msa@0E
\mathchardef\vDash="3\msa@0F
\mathchardef\twoheadrightarrow="3\msa@10
\mathchardef\twoheadleftarrow="3\msa@11
\mathchardef\leftleftarrows="3\msa@12
\mathchardef\rightrightarrows="3\msa@13
\mathchardef\upuparrows="3\msa@14
\mathchardef\downdownarrows="3\msa@15
\mathchardef\upharpoonright="3\msa@16

\mathchardef\downharpoonright="3\msa@17
\mathchardef\upharpoonleft="3\msa@18
\mathchardef\downharpoonleft="3\msa@19
\mathchardef\rightarrowtail="3\msa@1A
\mathchardef\leftarrowtail="3\msa@1B
\mathchardef\leftrightarrows="3\msa@1C
\mathchardef\rightleftarrows="3\msa@1D
\mathchardef\Lsh="3\msa@1E
\mathchardef\Rsh="3\msa@1F
\mathchardef\rightsquigarrow="3\msa@20
\mathchardef\leftrightsquigarrow="3\msa@21
\mathchardef\looparrowleft="3\msa@22
\mathchardef\looparrowright="3\msa@23
\mathchardef\circeq="3\msa@24
\mathchardef\succsim="3\msa@25
\mathchardef\gtrsim="3\msa@26
\mathchardef\gtrapprox="3\msa@27
\mathchardef\multimap="3\msa@28
\mathchardef\therefore="3\msa@29
\mathchardef\because="3\msa@2A
\mathchardef\doteqdot="3\msa@2B

\mathchardef\triangleq="3\msa@2C
\mathchardef\precsim="3\msa@2D
\mathchardef\lesssim="3\msa@2E
\mathchardef\lessapprox="3\msa@2F
\mathchardef\eqslantless="3\msa@30
\mathchardef\eqslantgtr="3\msa@31
\mathchardef\curlyeqprec="3\msa@32
\mathchardef\curlyeqsucc="3\msa@33
\mathchardef\preccurlyeq="3\msa@34
\mathchardef\leqq="3\msa@35
\mathchardef\leqslant="3\msa@36
\mathchardef\lessgtr="3\msa@37
\mathchardef\backprime="0\msa@38
\mathchardef\risingdotseq="3\msa@3A
\mathchardef\fallingdotseq="3\msa@3B
\mathchardef\succcurlyeq="3\msa@3C
\mathchardef\geqq="3\msa@3D
\mathchardef\geqslant="3\msa@3E
\mathchardef\gtrless="3\msa@3F
\mathchardef\sqsubset="3\msa@40
\mathchardef\sqsupset="3\msa@41
\mathchardef\trianglerighteq="3\msa@44
\mathchardef\trianglelefteq="3\msa@45
\mathchardef\bigstar="0\msa@46
\mathchardef\between="3\msa@47
\mathchardef\blacktriangledown="0\msa@48
\mathchardef\blacktriangleright="3\msa@49
\mathchardef\blacktriangleleft="3\msa@4A
\mathchardef\blacktriangle="0\msa@4E
\mathchardef\triangledown="0\msa@4F
\mathchardef\eqcirc="3\msa@50
\mathchardef\lesseqgtr="3\msa@51
\mathchardef\gtreqless="3\msa@52
\mathchardef\lesseqqgtr="3\msa@53
\mathchardef\gtreqqless="3\msa@54
\mathchardef\Rrightarrow="3\msa@56
\mathchardef\Lleftarrow="3\msa@57
\mathchardef\veebar="2\msa@59
\mathchardef\barwedge="2\msa@5A
\mathchardef\doublebarwedge="2\msa@5B
\mathchardef\angle="0\msa@5C
\mathchardef\measuredangle="0\msa@5D
\mathchardef\sphericalangle="0\msa@5E
\mathchardef\varpropto="3\msa@5F
\mathchardef\smallsmile="3\msa@60
\mathchardef\smallfrown="3\msa@61
\mathchardef\Subset="3\msa@62
\mathchardef\Supset="3\msa@63
\mathchardef\Cup="2\msa@64

\mathchardef\Cap="2\msa@65

\mathchardef\curlywedge="2\msa@66
\mathchardef\curlyvee="2\msa@67
\mathchardef\leftthreetimes="2\msa@68
\mathchardef\rightthreetimes="2\msa@69
\mathchardef\subseteqq="3\msa@6A
\mathchardef\supseteqq="3\msa@6B
\mathchardef\bumpeq="3\msa@6C
\mathchardef\Bumpeq="3\msa@6D
\mathchardef\lll="3\msa@6E

\mathchardef\ggg="3\msa@6F

\mathchardef\circledS="0\msa@73
\mathchardef\pitchfork="3\msa@74
\mathchardef\dotplus="2\msa@75
\mathchardef\backsim="3\msa@76
\mathchardef\backsimeq="3\msa@77
\mathchardef\complement="0\msa@7B
\mathchardef\intercal="2\msa@7C
\mathchardef\circledcirc="2\msa@7D
\mathchardef\circledast="2\msa@7E
\mathchardef\circleddash="2\msa@7F
\def\ulcorner{\delimiter"4\msa@70\msa@70 }
\def\urcorner{\delimiter"5\msa@71\msa@71 }
\def\llcorner{\delimiter"4\msa@78\msa@78 }
\def\lrcorner{\delimiter"5\msa@79\msa@79 }
\def\yen{\mathhexbox\msa@55 }
\def\checkmark{\mathhexbox\msa@58 }
\def\circledR{\mathhexbox\msa@72 }
\def\maltese{\mathhexbox\msa@7A }
\mathchardef\lvertneqq="3\msb@00
\mathchardef\gvertneqq="3\msb@01
\mathchardef\nleq="3\msb@02
\mathchardef\ngeq="3\msb@03
\mathchardef\nless="3\msb@04
\mathchardef\ngtr="3\msb@05
\mathchardef\nprec="3\msb@06
\mathchardef\nsucc="3\msb@07
\mathchardef\lneqq="3\msb@08
\mathchardef\gneqq="3\msb@09
\mathchardef\nleqslant="3\msb@0A
\mathchardef\ngeqslant="3\msb@0B
\mathchardef\lneq="3\msb@0C
\mathchardef\gneq="3\msb@0D
\mathchardef\npreceq="3\msb@0E
\mathchardef\nsucceq="3\msb@0F
\mathchardef\precnsim="3\msb@10
\mathchardef\succnsim="3\msb@11
\mathchardef\lnsim="3\msb@12
\mathchardef\gnsim="3\msb@13
\mathchardef\nleqq="3\msb@14
\mathchardef\ngeqq="3\msb@15
\mathchardef\precneqq="3\msb@16
\mathchardef\succneqq="3\msb@17
\mathchardef\precnapprox="3\msb@18
\mathchardef\succnapprox="3\msb@19
\mathchardef\lnapprox="3\msb@1A
\mathchardef\gnapprox="3\msb@1B
\mathchardef\nsim="3\msb@1C
\mathchardef\napprox="3\msb@1D
\mathchardef\nsubseteqq="3\msb@22
\mathchardef\nsupseteqq="3\msb@23
\mathchardef\subsetneqq="3\msb@24
\mathchardef\supsetneqq="3\msb@25
\mathchardef\subsetneq="3\msb@28
\mathchardef\supsetneq="3\msb@29
\mathchardef\nsubseteq="3\msb@2A
\mathchardef\nsupseteq="3\msb@2B
\mathchardef\nparallel="3\msb@2C
\mathchardef\nmid="3\msb@2D
\mathchardef\nshortmid="3\msb@2E
\mathchardef\nshortparallel="3\msb@2F
\mathchardef\nvdash="3\msb@30
\mathchardef\nVdash="3\msb@31
\mathchardef\nvDash="3\msb@32
\mathchardef\nVDash="3\msb@33
\mathchardef\ntrianglerighteq="3\msb@34
\mathchardef\ntrianglelefteq="3\msb@35
\mathchardef\ntriangleleft="3\msb@36
\mathchardef\ntriangleright="3\msb@37
\mathchardef\nleftarrow="3\msb@38
\mathchardef\nrightarrow="3\msb@39
\mathchardef\nLeftarrow="3\msb@3A
\mathchardef\nRightarrow="3\msb@3B
\mathchardef\nLeftrightarrow="3\msb@3C
\mathchardef\nleftrightarrow="3\msb@3D
\mathchardef\divideontimes="2\msb@3E
\mathchardef\varnothing="0\msb@3F
\mathchardef\nexists="0\msb@40
\mathchardef\mho="0\msb@66
\mathchardef\thorn="0\msb@67
\mathchardef\beth="0\msb@69
\mathchardef\gimel="0\msb@6A
\mathchardef\daleth="0\msb@6B
\mathchardef\lessdot="3\msb@6C
\mathchardef\gtrdot="3\msb@6D
\mathchardef\ltimes="2\msb@6E
\mathchardef\rtimes="2\msb@6F
\mathchardef\shortmid="3\msb@70
\mathchardef\shortparallel="3\msb@71
\mathchardef\smallsetminus="2\msb@72
\mathchardef\thicksim="3\msb@73
\mathchardef\thickapprox="3\msb@74
\mathchardef\approxeq="3\msb@75
\mathchardef\succapprox="3\msb@76
\mathchardef\precapprox="3\msb@77
\mathchardef\curvearrowleft="3\msb@78
\mathchardef\curvearrowright="3\msb@79
\mathchardef\digamma="0\msb@7A
\mathchardef\varkappa="0\msb@7B
\mathchardef\hslash="0\msb@7D
\mathchardef\hbar="0\msb@7E
\mathchardef\backepsilon="3\msb@7F
\def\Bbb{\ifmmode\let\next\Bbb@\else
 \def\next{\errmessage{Use \string\Bbb\space only in math mode}}\fi\next}
\def\Bbb@#1{{\Bbb@@{#1}}}
\def\Bbb@@#1{\fam\msbfam#1}

\catcode`\@=\active

\def\be{\begin{equation}}
\def\ee{\end{equation}}
\def\ba{\begin{eqnarray}}
\def\ea{\end{eqnarray}}

\def\v#1{\vert #1 \rangle}

\def\sp#1#2{\langle #1 \vert #2 \rangle}
\def\pa#1{\partial^#1}

\def\pax{\partial}
\def\paxi{\int_0^{\cdot}}
\def\ga{g_{\alpha}}

\def\intd{\int_{\cdot}^{\cdot}}
\def\fun{{\cal A}}
\def\bx{{\bf x}}
\def\intinf#1{{\int^{#1\infty}_{-#1\infty}}}

\def\CR{\hbox{{$\cal R$}}}

\def\C{{\Bbb C}}
\def\Z{{\Bbb Z}}

\def\vect{{\bf t}}\def\vecv{{\bf v}}
\def\vecx{{\bf x}}
\def\vecy{{\bf y}}
\def\vecw{{\bf w}}

\def\<{\langle}
\def\>{\rangle}
\def\del{{\partial}}
\def\lform{\hbox{$\sqcup$}\llap{\hbox{$\sqcap$}}}
\def\h{{{1\over2}}}
\def\eps{{\epsilon}}

\def\tens{\mathop{\otimes}}
\def\ra{{\triangleleft}}

\def\coev{{\rm coev}}

\def\id{{\rm id}}

\def\proof{\goodbreak\noindent{\bf Proof\quad}}

\def\endproof{{\ $\lform$}\bigskip }

\def\und#1{{\underline {#1}}}

\def\note#1{}

\def\equad{\kern -1.7em}

\def\eqn#1#2{\begin{equation}#2\label{#1}\end{equation}}
\def\cmath#1{\[\begin{array}{c} #1 \end{array}\]}

\def\ceqn#1#2{\begin{equation}\label{#1}
\begin{array}{c}#2\end{array}\end{equation}}
\def\align#1{\begin{eqnarray*}#1\end{eqnarray*}}
\def\alignn#1#2{\begin{eqnarray}\label{#1}#2
\end{eqnarray}}
\def\vol{{\rm vol}}
\def\lvec#1{{\overleftarrow {#1}}}
\def\dila{{\varsigma}} % if you havent got the font, try another letter

\def\Zint{{Z}}
\def\fou{{\cal F}}

\documentstyle[12pt,epsf]{article}
\newtheorem{lemma}{Lemma}[section]
\newtheorem{propos}[lemma]{Proposition}

\newtheorem{theorem}[lemma]{Theorem}

\begin{document}\baselineskip 22pt

{\ }\hskip 4.3in DAMTP/94-7/rev.  %put here preprint number
\vspace{.5in}

\begin{center}\baselineskip 13pt
 {\LARGE Algebraic $q$-Integration and Fourier Theory on Quantum and Braided
Spaces}\\
{\ }\\
Achim Kempf\footnote{supported by Studienstiftung des deutschen Volkes,
BASF-fellow}\\ {\ }\\ \&\\{\ } \\
Shahn Majid\footnote{Royal Society University Research Fellow and Fellow of
Pembroke College, Cambridge}
\\ {\ }\\
Department of Applied Mathematics\\
\& Theoretical Physics\\ University of Cambridge\\ Cambridge CB3 9EW, U.K.\\

\vspace{10pt}

January, 1994\\

\vspace{10pt}
{\bf Abstract}
\end{center}
\begin{quote}
\baselineskip 13pt
We introduce an algebraic theory of integration on quantum planes and other
braided spaces. In the one dimensional case we obtain a novel picture of the
Jackson $q$-integral as indefinite integration on the braided group of
functions in one variable $x$. Here
$x$ is treated with braid statistics $q$ rather than the usual bosonic or
Grassmann ones. We show that the definite integral $\intinf x$ can also be
evaluated algebraically as multiples of the integral of a $q$-Gau{\ss}ian, with
$x$ remaining as a bosonic scaling variable associated with the
$q$-deformation. Further composing our algebraic integration with a
representation then leads to ordinary numbers for the integral. We also use our
integration to develop a full theory of $q$-Fourier transformation $\fou$. We
use the braided addition $\Delta x=x\tens 1+1\tens x$ and braided-antipode $S$
to define a convolution product, and prove a convolution theorem. We prove also
that $\fou^2=S$. We prove the analogous results on any braided group, including
integration and Fourier transformation on quantum planes associated to general
R-matrices, including $q$-Euclidean and $q$-Minkowski spaces.
\end{quote}
\baselineskip 21pt
PACS Numbers:  03.65.Fd, 02.30.Nw, 02.30.Cj, 02.10.Rn, 11.30.-j
\newline Short title: $q$-Integration and Fouirer Theory on Quantum Spaces

\section{Introduction}

This work continues a series of papers
\cite{Ma:exa}\cite{Ma:lin}\cite{Ma:poi}\cite{Ma:fre}
\cite{Ma:lie}\cite{Mey:new}\cite{Ma:euc} in which is developed a systematic
approach to $q$-deforming physics based on the idea that the geometrical
co-ordinates should have braid statistics. Such co-ordinates
are a generalisation of usual Bose or Fermi ones with $\pm1 $ replaced by a
parameter $q$, or more generally by an R-matrix. This gives us a generalisation
of super-geometry as some kind of `braided-geometry'.  Here
\cite{Ma:exa}\cite{Ma:lin}\cite{Ma:poi} introduced the basic notions of
braided-matrices, braided vectors, their addition law and their covariance
properties under a background quantum group, i.e. a complete covariant and
braided linear algebra. \cite{Ma:fre} introduced the notion of braided
differentiation on braided vectors spaces as an infinitesimal translation, and
also the general braided exponential map as its eigenfunctions. We refer to
\cite{Ma:introp} for a review of the whole braided approach. Our goal now is to
take a step towards completing this programme by providing also the beginnings
of a general theory of braided or $q$-deformed integration and Fourier
transformation. We treat the one-dimensional case in complete detail and some
aspects of the general theory. The standard quantum plane in $n$-dimensions is
covered completely as well. The algebras for $q$-Euclidean and $q$-Minkowski
space are known and our approach applies to these also.

The general programme of $q$-deforming physics is both a classical subject in
the context of $q$-special functions, see e.g.\cite{Exton}\cite{And:ser}, and a
currently popular one in the context of quantum groups and non-commutative
geometry. While compatible with some of this previous work, the braided
approach above differs in the following fundamental way: the $q$-deformation
which we introduce is not directly into the co-ordinate algebra of the system
but rather into non-commutativity of the tensor product algebra $\tens_q$ of
two independent copies of the system.

All of this is visible even in one dimension, where we just have one variable
$x$ say to which we apply $q$-derivatives etc. Usually, one would consider $x$
at least in this case as a number since it always commutes with itself. In the
braided approach however, this  generator $x$ must remain an operator or
abstract generator because it will not commute with other copies of itself. It
was shown in \cite{Ma:poi} how such braid statistics lead at once to the usual
$q$-derivative $\del_q$ and $q$-exponential. The key idea is that functions in
one variable form a braided-Hopf algebra called the
`braided-line'\cite{Ma:csta}. This consists of the functions $\C[x]$ equipped
with a braided coaddition $\Delta:\C[x]\to \C[x]\und\tens\C[y]$ sending $f(x)$
to $\Delta f=f(x+y)$, but where the two copies $q$-commute according to
$qyx=xy$. We can then define
\eqn{qdif}{ (\del_q
f)(y)=\left(x^{-1}(f(x+y)-f(y))\right)|_{x=0}={f(y)-f(qy)\over (1-q)y}}
which is the standard $q$-derivative.  There is also a `counit' and
braided-antipode
\eqn{ant}{ \eps f=f(0),\quad S x^n=(-1)^n q^{n(n-1)\over 2} x^n}
for evaluation at zero and $q$-subtraction. The braided antipode $S$ is one of
the important new ingredients in $q$-analysis provided by the theory of braided
groups.  The $q$-exponential is likewise understood in braided group theory as
a braided-multiplicative element:
\eqn{qexp}{ \Delta e_q^{x\lambda}=e_q^{x\lambda}\tens e_q^{x\lambda},\quad
Se_q^{x\lambda}=e_{q^{-1}}^{-x\lambda}}
Moreover, it is this new braided point of view on $q$-differentials and
$q$-exponentials which generalises at once to arbitrary dimensions and
arbitrary R-matrices\cite{Ma:fre}.

We want to apply these same techniques now to obtain an understanding of the
celebrated Jackson $q$-integral \cite{Jac:int}  $\int_0^a d_qx$, as well as to
generalise it to higher dimensions. The point of view that we are led to is a
fairly radical one due to the fact that the variable $x$ is for us a braided
co-ordinate and hence not a $\C$-number as it would be in the usual point of
view. We consider the limits of the Jackson integral as braided variables and
hence the integral itself as operators of indefinite integration $\int^{x}_0$
and $\int^y_x=\int_0^y-\int_0^x$. Then one finds
\alignn{intx+y}{\int^{x+y}_0 f\equad &&=\int^y_0 f+\int^x_0 f((\ )+y),\quad{\rm
if}\quad y[0,x]=q[0,x]y\\
&&=\int^x_0 f+\int^y_0 f(x+(\ )),\quad {\rm if}\quad  [0,y]x=qx[0,y]}
where the braid statistics $y[0,x]=q[0,x]y$  means to assume $yx=qxy$ when
computing $\int^{x+y}_0$ and also $yz=qzy$ in  $f(z+y)$ when computing
$\int^x_0$ with variable of integration $z$. Similarly for the second
indentity. These two identities are easily verified on monomials using the
well-known  $q$-binomial theorem. We see that we can exactly think of
$\int^y_0$ as built up by a Riemann sum provided the small interval $[0,x]$
being added is treated with braid statistics relative to the points in the
existing sum. We arrive at exactly an inverse to $q$-differentiation from the
point of view of (\ref{qdif}). Indeed, supposing $x$ `small' we have
\[  \int^{x+y}_0 f-\int^y_0 f=\int^x_0 f((\ )+y)\approx xf(y)\]
as in the usual theory of integration. From these identities, one also has
\ceqn{intxy+}{\int^{z+y}_{z+x} f=\int_x^y f(z+(\ ))\quad {\rm if}\quad
[x,y]z=qz[x,y]\\
\int^{y+z}_{x+z} f=\int_x^y f((\ )+z)\quad {\rm if}\quad z[x,y]=q[x,y]z}
which is global translation-covariance in our approach. One can think of the
functions on the braided line as an infinite braided tensor product of $\C$
with the tensor factors ordered lexicographically on a line cf.\cite[Sec.
6]{Ma:introp}.
Then the braid statistics are $yx=qxy$ whenever $y>x$.

This braided point of view requires us to think of integration as an operator
rather than to evaluate it as a number. In Section~2 we take this further and
write the integration operator as a powerseries in the $q$-Heisenberg algebra
generated by the operator ${\bf x}$ of multiplication by the co-ordinate
function, and $\pax=\del_{q^2}$. The idea is to work abstractly in this
algebra, with the question of convergence or not (i.e. of integrability of a
function) becoming a property of any representation in which we might view it.
We take this line also for integration $\int_0^{x\infty}$, which we define by
infinitely scaling up $\int_0^x$. So our braided $x$ remains as a scale
parameter with its braid-statistics `diluted' to the extent that in the limit,
it behaves bosonically. As before, limits are to be evaluated only in
representations.

Section~3 is  devoted to our first such family of representations, labelled by
a parameter $c$ and motivated by a quantum-mechanical picture of the
$q$-Heisenberg algebra. We show that in these representations, we recover the
formulae and values given by Jackson integration in its usual form. Also within
these representations there is a reasonable notion of `points' as eigenvectors
of the position operator, and of `integration regions' as sub spaces of the
representation space. Our representations also provide a  trace formula for
Jackson integration. So far, we have developed these representations only in
the 1-dimensional case.

Section~4 develops a second evaluation procedure which is like a representation
in that it yields numbers, as multiples of a single undetermined integral which
remains as a bosonic element in our algebra. This is done by integrating with
reference to a suitable $q$-Gau{\ss}ian function $g$, which we also introduce.
Our approach here is similar in spirit to the example of
\cite{BauFlo:pat}\cite{Kem:sym} for integration over the entire complex plane,
but somewhat different. We give a condition for integrability of a function
according to our scheme. There are plenty of integrable functions in this
sense, for example polynomials times our reference Gau{\ss}ian. In general
terms, our algebraic approach is analogous to the treatment of path integration
in physics where all integrals are computed relative to a reference
Gau{\ss}ian.

In Section~5 we construct integration on general $n$-dimensional quantum
planes, in the same spirit as our treatment above. In the simple case of the
standard $SL_q(n)$ quantum planes we show that this can effectively be computed
by iterating the 1-dimensional integration with integration operators
$\int^{x_i}_0$, cf. the iterated integrals in \cite{KorVak:spa}.
In general however, these operators from the quantum
co-ordinate ring to itself are somewhat more complicated. They have
$q$-commutation relations  like the co-ordinates $x_i$. We then give the
Gau{\ss}ian approach in the general R-matrix setting of quantum and braided
spaces as introduced in \cite{Ma:poi}. The main theorem is Theorem~5.1 and
expresses the Gau{\ss}ian-weighted integral of monomials in terms of an
interesting factorisation property. For example,
\[ {\int x_{i_1}x_{i_2}x_{i_3}x_{i_4} g\over\int g}= {\int x_{i_1}x_{i_2}
g\over\int g} {\int x_{i_3}x_{i_4} g\over\int g}+ \lambda^{2}{\int x_{i_1}x_{a}
g\over\int g} {\int x_{i_2}x_{b} g\over\int
g}R^a{}_{i_3}{}^b{}_{i_4}+\lambda^{4}{\int x_{i_1}x_{a} g\over\int g} {\int
x_{b}x_{c} g\over\int g}R^b{}_{i_2}{}^a{}_d R^c{}_{i_3}{}^d{}_{i_4}\]
where $g$ is the Gau{\ss}ian and $R$ the general R-matrix. $\lambda$ is a
normalisation constant. We compute the integrals on $q$-Euclidean and
$q$-Minkowski spaces as examples.

Finally, in Section~6 we develop an easy application of our integration theory,
making use of the translation invariance with respect to $\Delta$ for
integration on the braided-line and other braided spaces. Namely, we give a
fairly complete theory of $q$-Fourier transforms on such spaces. While
$q$-exponentials, and also $q$-sine and $q$-cosine are known\cite{Exton}, the
theory that we describe would not have been possible before even in the one
dimensional case, as it depends for its meaningfulness on the notion of
braided-groups with $\Delta$ and antipode (\ref{ant}) as introduced only in the
last few years by the second author.

We do the general case first using the diagrammatic techniques associated with
braided groups\cite{Ma:introp}, i.e. we develop a kind of braided-calculus for
Fourier transforms which works for any braided group. Again, it is algebraic
and takes a concrete form with questions of convergence arising only in a
realisation. In this setting, we have a convolution theorem, $\delta$-functions
and $\fou^2=S$. We then pass to the 1-dimensional case and the general
multidimensional R-matrix case.

While this work was being written-up, we received a preprint\cite{ChrZum:tra}
where the braided-translation ideas introduced in \cite{Ma:poi}\cite{Ma:fre}
were also used to propose the possibility of a $q$-Fourier transform in the
Hecke case. On the other hand,
these authors use a quite different proposal for translation-invariant
integration, which, in one dimension, is just ordinary integration. Hence their
corresponding Fourier transform is quite different from our proposal based on
Jackson integration and its generalisation to $n$-dimensions.

We would also like to note two recent papers \cite{HebWei:fre}\cite{Fio:sym},
which
were pointed out to us, where translation invariant integration on
$q$-Euclidean space was considered in the context of $SO_q(N)$-invariant
quantum mechanics. The second of these used an explicit Gau{\ss}ian approach,
though not in our general $R$-matrix setting.

\baselineskip 15pt
\tableofcontents
\baselineskip 22pt

\section{Algebraic integration}

In this section we develop an algebraic point of view on ordinary
$q$-integration in one variable $x$ as, quite literally, the inverse operator
to $q$-differentiation. We study such an operator $\int_0^x$ and develop its
elementary properties. After that, we obtain integration to $\infty$ by scaling
the indefinite one. We also explain the issues relevant to the $n$-dimensional
case to come later.

\subsection{Indefinite integration}

Let us consider the algebra $\cal{Q}$ generated by a multiplication
operator ${\bf x}$ and a differentiation operator $\pax$ obeying
the commutation relations:
\be
\pax {\bf x} - q^2 {\bf x} \pax = 1 \qquad \mbox{ with } \qquad q^2<1
\label{cr1dim}
\ee
The algebra is naturally represented on the algebra
$\fun $ of polynomials or other suitable
functions in $x$, with $\del$ represented as in (\ref{qdif}) except
that we use conventions with $q^2$ rather than $q$. It will later be
instructive to consider a different representation also.

Unlike in ordinary calculus, which is the case $q=1$, it is now
possible to identify an operator $\int_0^.$
or $\partial^{-1}$ as an element
of the algebra $\cal{Q}$. We start by formally writing:
\be
\paxi = {\partial}^{-1} = {\bf x} {\bf x}^{-1} \paxi = {\bf x} (\pax {\bf
x})^{-1} =
 {\bf x} \frac{1-q^2}{1- (1-(1-q^2)\pax {\bf x})}
\ee
{}From the action of the term $ (1-(1-q^2)\pax {\bf x})$:
\be
(1-(1-q^2)\pax {\bf x}).x^r = q^{2(r+1)} x^{r}
\ee
we see that it is diagonal in the basis of monomials and has
the eigenvalues $q^{r+1}$ with $r=0,1,2,...$.
Thus we can treat this operator
 like a number of absolute
value smaller than $1$ and the expansion
yields:
\be
\paxi = (1-q^2) {\bf x} \sum_{n=0}^{\infty} (1-(1-q^2)\pax {\bf x})^n
\label{dinv}
\ee
We will see in Section~5 that these arguments  also go through
for the differential calculus on the $SL_q(n)$ and other quantum planes. In
this case, however, the function algebra $\fun $ generated by the co-ordinate
functions on the quantum planes will be noncommutative.

To complete our operator picture, note that ordinary integration
$$
f(t) \rightarrow \int_{t_1}^{t_2} dt^{\prime} f(t^{\prime})
$$
is really a mapping from a space of functions of one variable to a space of
functions of two variables, namely the
upper and the lower limits of the integration interval. In our case, we define
analogously
\be
\intd : \fun  \rightarrow \fun  \otimes \fun,\quad \intd := \paxi \otimes 1 - 1
\otimes \paxi.
\label{interval}
\ee
One readily checks the desired properties
\eqn{dinta}{(\pax \otimes id) \intd = id \otimes 1}
\eqn{dintb}{(id \otimes \pax) \intd  = -1 \otimes id}
\eqn{dintc}{\intd \pax = id \otimes 1 - 1 \otimes id}
and the global translation properties
\eqn{inttrans}{\Delta\circ\paxi=1\tens\paxi+(\paxi\tens\id)\circ\Delta
=\paxi\tens 1+(\id\tens\paxi)\circ\Delta}
\eqn{intdtrans}{\beta_L\circ\intd=(\id\tens\intd)\circ\Delta,\quad
\beta_R\circ\intd=(\intd\tens\id)\circ\Delta.}
Here $\Delta$ is the coaddition for the braided line, which we view as a left
or right coaction of $\fun$ on itself. Equation (\ref{intdtrans}) says that
$\intd$ is an intertwiner between these coactions and the corresponding induced
tensor product coactions $\beta_L,\beta_R$ on $\fun\tens\fun$. This is the
operator description of our observations (\ref{intx+y})--(\ref{intxy+}) from
the Introduction.

\subsection{Scaling and infinity}

Let us now develop a simple method for
integration `over the whole space'. We will not aim to extract the
 information of what the value of the integral is from the
`values' the function has in the integration interval. Instead
we will obtain the integral in terms of the coefficients
of the power series expansion of the function, which is
the form in which our function algebra $\fun $ is
naturally given in the quantum group framework.
We later use this formalism in the $n$-dimensional case also.

Consider the `scaling' operator:
\be
L := 1 - (1-q^2){\bf x}\pax \qquad \in \cal{Q}
\label{defL}
\ee
Its action on the monomials in $\fun $ is
\be
L.x^r = q^{2r} x^r
\ee
Thus, it scales the functions in $\fun $:
\be
L.f(x) = f(q^2 x)
\ee
The inverse can also be found in $\cal{Q}$:
\be
L^{-1} = \frac{1}{1 - (1-q^2){\bf x}\pax} = \sum_{s=0}^{\infty}
((1-q^2) {\bf x}\pax)^s
\ee
As we saw in Section 2.1, the integral of a function should be considered
a function in its integration limits. We now have an algebraic tool at hand to
scale these
limits. In particular one readily checks that
\be
L^{-r} \paxi = {\bf x} (1-q^2) \sum_{n = -r}^{\infty}
(1-(1-q^2)\pax {\bf x})^n
\ee
leading us to define
\be
\int_0^{x\infty} := \lim_{r\rightarrow \infty}
L^{-r} \paxi = {\bf x}(1-q^2) \sum_{n = -\infty }^{\infty}
(1-(1-q^2)\pax {\bf x})^n.
\ee
For each finite $r$, this operator remains a map from
the function space $\fun $ into $\fun $. In the generic case then the integral
is the limit of
a power series in the generator. So, although we are integrating over the whole
space, the integral
\be
I(f) := \int_0^{x\infty} f
\ee
of a function $f$ is still an expression
in the variable $x$. In ordinary calculus this would be a constant function,
while in our case we can
say at least that it is invariant under the
scaling operation
\eqn{scalinv}{L^{\pm 1} I(f) = I(f)}
This is obvious from its definition. In the braided picture it means that
$I(f)$ is bosonic in the sense that $yI(f)=I(f)y$ even if $y$ $q$-commutes with
$x$.

This point of view on global integration extends to the $n$-dimensional case
also. We will see that one has similar scaling operators $L_i$ in the case of
$SL_q(n)$ quantum planes and can use them to define global integration by
infinite scaling in the same manner as above. Now the various
co-ordinates $x_i$ are noncommuting even among themselves and both the
indefinite integral and the global integral of a function in our noncommutative
algebra remain as functions of these noncommuting variables. In this case it is
even more clear that
it would be naive to try to `evaluate' such functions at some
finite `point' by simply putting in numbers for the noncommutative
generators. The same function, rearranged using only the
commutation relations would have another value at the same `point'.
In general one does not have many nontrivial algebra
homomorphisms from the noncommutative function algebra
to the commutative ground field.

However, for physical applications it is of course very desirable
 to obtain actual numbers from the integration. We see
at least two strategies to achieve this:

1. One could find other concrete representations and try to define
 `quasipositions' as eigenvectors of the position operators.
Due to their noncommutativity they will not be simultaneously
 diagonalisable, even if the $*$-structure is such that they are
 symmetric. Integration regions then appear as sub-Hilbert
spaces of the representation space. We will demonstrate the
 idea in the simple one dimensional case in the next section.

2. For integration over the whole space one can use the
following technique: If $g$ is a
suitable function that is integrable over the whole
space, it is possible to generate a space of integrable
functions from it. The integral of $g$ may in general be a
noncommutative expression. However, the integrals
of all other integrable functions are simply
multiples of the integral of $g$. The key point is that
for appropriately chosen $g$ all boundary terms of partial
 integrations vanish, so that no new noncommutative expressions
other than the integral of $g$ can appear. We will develop
this technique in Section 4 for the one dimensional case, and
we will then extend it to the true noncommutative case in
Section 5.

\section{Quasipositions representation}

In the preceding section we represented the algebra of
${\bf x}$ and $\pax$ as operators on the function space $\fun $. Following the
first strategy explained above,
we consider another representation of the algebra $\cal{Q}$ spanned now by the
eigenvectors of the position
operator ${\bf x}$. Note that (\ref{cr1dim}) allows us to choose ${\bf x}$ to
be symmetric. However,
unlike in ordinary calculus, $\pax$ will then no longer be
antisymmetric.

\subsection{Construction of the representation}
We construct the representation by starting with a normalised eigenvector
of the position operator to the eigenvalue $c$:
\be
{\bf x} \v{v_c} = c \v{v_c} \mbox{ \qquad with \qquad} \sp{v_c}{v_c}=1
\ee
In $\cal{Q}$ we have
\be
{\bf x} (1-(1-q^2)\partial {\bf x}) = q^{-2} (1-(1-q^2)\partial  {\bf x}) {\bf
x}
\ee
Thus the normalised vectors
\be
\v{v_{c q^{-2r}}} := N(r) (1-(1-q^2)\partial  {\bf x})^r \v{v_c}
\label{states}
\ee
are eigenvectors of ${\bf x}$ with the eigenvalues $c q^{-2r}$.
There is also the inverse operator
\be
\frac{1}{1-(1-q^2)\partial  {\bf x}} = \sum_{m=0}^{\infty} ((1-q^2)\partial
{\bf x})^m
\ee
since the eigenvalues of $(1-q^2)\partial  {\bf x}$ are
$1-q^{2r}$ i.e. smaller than $1$.
We can thus let $r$ in (\ref{states}) run through all integers.
Evidently, the scale $c$  labels the representation.

We consider the eigenvectors $\vert v_{cq^{2r}} \rangle$ of ${\bf x}$
as the `quasipoints'. We can then think of general vectors as
'functions' on the set of quasipoints, i.e. with values as given by
evaluation against quasipoint vectors. Thus
\be
\v{f} = \sum_{r} f_r \v{v_{cq^{2r}}}
\quad \mbox{ has values } \quad
\langle v_{cq^{2r}} \v{f} = f_r
\label{fctns1}
\ee
The position operator ${\bf x}$ then appears as a matrix
\be
x = \sum_{r,s} \vert v_{cq^{2r}}\rangle cq^{2r} \delta_{r,s}
\langle v_{cq^{2s}}\vert
\ee
In the representation $\fun $ that we had considered so far, the functions are
polynomials or power series in the variable $x$. Each such function $f(x)$ can
now uniquely be identified with a vector in the quasipoint representation.
Let us first consider the constant polynomial $1 \in \fun $. It is
identified with that vector $\v{1}$ in the
quasipoint representation which
maps every quasipoint onto the number $1$, i.e.
\be
\v{1} = \sum_{t=-\infty}^{+\infty} \v{v_{cq^{2t}}}
\ee
Obviously at each quasipoint $\v{v_{cq^{2t}}}$ the value of
$\v{1}$ is $\sp{v_{cq^{2r}}}{1} = 1$.

Consider now an arbitrary polynomial or power series $f(x) \in \fun $. It is
identified with a vector in the quasipoint representation which is
obtained as follows. We simply let $f(\bx)$ act, as a polynomial or power
series
in the position operator $\bx$, on the constant function $\v{1}$
\be
\v{f} := f(\bx) \v{1}
\label{fctns2}
\ee
On the other hand, just like in ordinary calculus, of course not
every function from (quasi-) points to numbers can
also be expressed as a polynomial or power series.

For the matrix representation of the differentiation operator
\be
%\partial \v{f} = \sum_{r,s} d_{r,s} f_s  \v{v_{cq^{2r}}}
\partial = \sum_{r,s} \vert v_{cq^{2r}}\rangle
d_{r,s}
\langle v_{cq^{2s}}\vert
\ee
equation (\ref{cr1dim}) yields
\be
d_{r,s} = \delta_{r,s} (cq^{2r}-cq^{2(r+1)})^{-1} + a(r) \delta_{r+1,s}
\label{wa}
\ee
Thus the commutation relation does not determine the differentiation
operator completely. It is however fixed from the requirement that
the differential of a constant function still vanishes: Putting
equation (\ref{wa}) into
\be
\partial \v{1} = 0
\ee
fixes $a(r)$ so that the matrix elements of the differentiation operator are
\be
d_{r,s} = (\delta_{r,s}-\delta_{r+1,s}) (cq^{2r}(1-q^2))^{-1}.
\ee
As we said, even for $q<1$ we can keep the postion operator
symmetric $x^{\dagger} = x$.
However, from equation (\ref{cr1dim}) it is
clear that then $\partial$ will no longer be
antisymmetric, as it would be for $q=1$. Explicitely we have
\be
x^{\dagger} = x = \sum_{r=-\infty}^{+\infty}
\v{v_{cq^{2r}}} cq^{2r} \langle v_{cq^{2r}} \vert
\ee
\be
\partial = \sum_{r,s} \v{v_{cq^{2r}}}
(\delta_{r,s}-\delta_{r+1,s}) (cq^{2r}(1-q^2))^{-1}
\langle {v_{cq^{2s}}}\vert .
\ee
The hermitean conjugate of $\partial$ is thus
\be
\partial^{\dagger} = \sum_{r,s} \v{v_{cq^{2s}}}
(\delta_{r,s}-\delta_{r+1,s}) (cq^{2r}(1-q^2))^{-1}
\langle {v_{cq^{2r}}}\vert .
\ee
In order to get the normalisation constants we now calculate the step
operators and their adjoints explicitely as
\begin{eqnarray}
(1-(1-q^2)\partial x)^r &=&
\sum_{s} q^{2s} \v{v_{cq^{-2s}}}
\langle v_{cq^{-2(s+r)}} \vert \\
((1-(1-q^2)\partial x)^{\dagger})^r &=&
\sum_{s} q^{2s} \v{v_{cq^{-2(s+r)}}}
\langle v_{cq^{-2s}} \vert .
\end{eqnarray}
The normalisation condition and Eq.\ref{states} yield:
\begin{eqnarray}
1 &=& \sp{v_{cq^{2r}}}{v_{cq^{2r}}}\nonumber \\
 &=& N^2(r) \langle v_c\vert ((1-(1-q^2)\partial x)^r)^+
(1-(1-q^2)\partial x)^r \v{v_c}\nonumber\\
 &=& N^2(r) q^{4r} \sum_{s,t}  \langle v_c
\v{v_{cq^{-2(s+r)}}} \langle v_{cq^{-2s}}
\v{v_{cq^{-2t}}} \langle v_{cq^{-2(t+r)}} \v{v_c}\nonumber \\
 &=& N^2(r) q^{4r}
\end{eqnarray}
Thus the normalisation constants in (\ref{states}) are
\be
N(r) = q^{-2r}
\ee
The Hilbert space is of course isomorphic to $l^2$.

\subsection{The integral operator and the Jackson integral formula}

In order to calculate the matrix representation of the integral
 operator
\be
\paxi = \sum_{r,s} \vert v_{cq^{2r}}\rangle
(\paxi)_{r,s}
\langle v_{cq^{2s}}\vert
\ee
 we could represent the expression in (\ref{dinv}) or
 simply invert the matrix of $\partial$ to obtain:
%\be
%(d^{-1})_{r,s} - (d^{-1})_{r+1,s} = c q^{2r} (1-q^2) \delta_{r,s}
%\ee
\be
(\paxi)_{r,s} = c(1-q^2) \sum_{t=0}^{\infty} q^{2(r+t)}
\delta_{r+t,s}.
\ee
Thus
\be
\paxi = c(1-q^2)\sum_{r=-\infty,t=0}^{\infty} q^{2(r+t)}
\v{v_{cq^{2r}}}\langle v_{cq^{2(r+t)}}\vert
\ee
Using (\ref{interval}), the integral from the
`quasiposition' $cq^{2a}$ to the
quasiposition $cq^{2b}$ is:
\be
\int_{cq^{2a}}^{cq^{2b}}  f  =  \langle v_{cq^{2b}}\vert
\langle v_{cq^{2a}}\vert \intd \v{f}
=  \langle v_{cq^{2b}} \vert \paxi \v{f} -  \langle v_{cq^{2a}} \vert
\paxi \v{f}
\label{intab}
\ee
\be =  c(1-q^2)\sum_{t=0}
q^{2t} \left( q^{2b} \langle v_{cq^{2(b+t)}} \vert
 - q^{2a} \langle v_{cq^{2(a+t)}} \vert \right) \v{f}
\ee
which, using (\ref{fctns1}) can be brought into the form of the
Jackson integral:
\be
\int_{cq^{2a}}^{cq^{2b}}  f =
cq^{2b} (1-q^2)\sum_{t=0}^{\infty} q^{2t} f(cq^{2b} q^{2t})
-cq^{2a}(1-q^2)\sum_{t=0}^{\infty}q^{2t} f(cq^{2a}q^{2t})
\ee

\subsection{A new trace formula for the integral}
On the other hand we will now also obtain a new trace
formula: From (\ref{intab}) we get for $a \rightarrow \infty$ and
$b \rightarrow -\infty$:
\be
\int_0^{c\infty} f = c(1-q^2) \sum_{t= -\infty}^{\infty}
q^{2t} \langle v_{cq^{2t}}\v{f} =
(1-q^2) \sum_{t= -\infty}^{\infty}
\langle v_{cq^{2t}}\vert {\bf x} \v{f}.
\ee
Now using (\ref{fctns2}) we obtain
\be
\int_0^{c\infty}  f =
(1-q^2) \sum_{s,t= -\infty}^{\infty}
\langle v_{cq^{2t}}\vert {\bf x} f({\bf x}) \v{v_{cq^{2s}}}
\ee
Since ${\bf x}f({\bf x})$ is diagonal we get for all functions $f \in \fun $:
\eqn{trace}{
\int_0^{c\infty}  f =
(1-q^2) \mbox{ Trace}( {\bf x} f({\bf x}))}
This is now a basis independent formulation and
one may expect it to be extendable to the general
$n$ dimensional case. Finite integration regions then
simply mean to take the trace only over
 a finite dimensional sub Hilbert space of the representation
space. We have in mind a similar representation theoretic approach in the
$n$-dimensional case also, though we will not
develop this explicitly here.

\section{Induced integration}

In this section we develop our algebraic approach to global integration based
on Gau{\ss}ians, in the one-dimensional
case. The presense of a Gau{\ss}ian weight factor in the integrand allows
integration to be `turned back' into  differentiation and thereby computed
purely algebraically.

\subsection{Rapidly decreasing functions}

We begin with a simple classification of a function's
behaviour at infinity, which will also prove to be useful in
the $n$-dimensional noncommutative case.
Let us say that a function $f \in \fun $ is
vanishing at $\pm \infty$ if it obeys:
\be
\lim_{n\rightarrow \infty} L^{-n} f(\pm x) = 0.
\ee
Similarly we say that a function $f \in \fun $ is rapidly decreasing
at infinity if we have the even stronger condition:
\be
\lim_{n\rightarrow \infty} L^{-n} x^r f(\pm x) = 0 \qquad \qquad
\mbox{ for all } r = 0,1,2,...
\ee
An important example of a function that is rapidly decreasing
at infinity is the function $\ga $ defined by the differential equation:
\be
\pax \ga = - \alpha x \ga \mbox{ \qquad with } \alpha > 0
\label{gaussde}
\ee
It is of course the analogue of a Gau{{\ss}}ian
 function $e^{-\alpha x^2/2}$. However, although we can
calculate it as a vector in representation space, it is not
trivial to write it as a power series in the operator $x$ acting
on the constant function $\vert 1\rangle$. Let us check that it is rapidly
decreasing.
Using (\ref{defL}) and (\ref{gaussde}) we get
 for any  $u(x) := x^r \ga $:
\be
L. u(x) = u(q^2 x) = (1+\alpha (1-q^2) x^2) q^{2r} u(x)
\ee
Thus:
\be
u(x) = \prod_{n=0}^r (1+ \alpha (1-q^2) q^{-2n} x^2) \quad q^{2r}
 u(q^{-2r} x)
\ee
Since the product is strictly increasing and diverging as
$r\rightarrow \infty$, the function $u(x)$ vanishes:
\be
\lim_{r\rightarrow \infty} L^{-r} x^r \ga = 0
 \qquad \qquad
\mbox{ for all } r = 0,1,2,...
\ee
In suitable representations then, we will have the
rapid decrease property of $\ga$ explicitly (see the appendix).
Since $\ga $ is an even function it has the same behaviour
at $-\infty$. For the same reason, its integral from
$-x\infty$ to $+x\infty$ is simply twice the integral
from $0$ to $+x\infty$.

In the following we will
not actually need any details about its global integral
\be
I(\ga) := \int_{-x\infty}^{x\infty} \ga.
\ee
In fact, we show in the Appendix that its actual
value
\be
(1-q^2) \mbox{Trace}({\bf x} \ga)
\ee
is  a finite number. This number depends nontrivially
on the chosen representation or, physically speaking, on the
choice of the scale $c$ that appears with the $q$-deformation.
Note that the trace is basis independent, but representation
dependent.
\subsection{Induced integrals}

Keeping in mind that $I(\ga)$ is not a number, but
representation dependent, we will now aim at expressing the global
 integrals of
other functions as multiples of $I(\ga)$.

To this end we consider
\be
\int_{-x\infty}^{x\infty} \pax x^r \ga =
2 \lim_{n\rightarrow \infty} L^{-n} x^r \ga = 0
\ee
which vanishes because $\ga$ is rapidly decreasing.

Thus
\be
0 = \int_{-x\infty}^{x\infty} \pax x^r \ga =
\int_{-x\infty}^{x\infty}\left( [r] x^{r-1} \ga +
q^{2r} x^r \pax \ga\right)\ee
where $[r]={1-q^{2r}\over 1-q^2}$ is the usual $q$-integer.
Using (\ref{gaussde}) yields the recursion relation
\be
\int_{-x\infty}^{x\infty} \alpha q^{2r} x^{r+1} \pax \ga =
\int_{-x\infty}^{x\infty} [r] x^{r-1} \ga
\ee
and eventually
\be
\int_{-x\infty}^{x\infty} x^r \ga =
I(\ga) \frac{[r-1]!!}{\alpha^{r/2} q^{r^2/2}} \mbox{ \quad for
all $r$ even}
\label{aif1}
\ee
and 0 for $r$ odd.
Now it is not difficult to prove that the global integral
of the function $f \ga$, for any $f \in \fun $ can be
obtained simply from the action of the operator $P_{\alpha}
 \in \cal{Q}$:
\be
\int_{-x\infty}^{x\infty} f \ga =
I(\ga)
P_{\alpha}.f\vert_{x=0}
\ee
where
\be
P_{\alpha} := \sum_{r=0}^{\infty}
    \alpha^{-r} q^{-2r^2} \frac{{\partial}^{2r}_x}{[2r]!!}.
\ee
In particular, every polynomial $f \in \fun $ times the
Gau{\ss}ian is integrable i.e.
$ P_{\alpha}.f\vert_{x=0}$ is finite.

\subsection{Global integration formula}

We eventually aim at expressing the global
integration over $f$ alone. To this end let us `undo' the
multiplication with $\ga$ by multiplication with $\ga^{-1}$.
Actually $\ga^{-1}$ can be found in $\fun $ as follows.

With the ansatz
\be
\ga^{-1} := \sum_{r=0}^{\infty} {\ga}^{-1}_r x^r
\ee
follows from
\be
\pax {\ga}^{-1} \ga = 0
\label{gig}
\ee
and (\ref{gaussde}) that
\be
\ga^{-1}(x) = \sum_{r=0}^{\infty}
 \frac{\alpha^r q^{2(r^2 -r)}}{[2r]!!} x^{2r}
\ee
A simple ratio test proves that this is
convergent everywhere, i.e.
for all eigenvalues of $x$.
 We thus express the
global integral of an arbitrary function
 i.e. power series $h \in \fun $ as a
multiple of the global integral $I(\ga)$ of the Gau{\ss}ian:
\be
\int_{-x\infty}^{x\infty}  h(x) =
\lim_{n\rightarrow \infty}
\int_{-x\infty}^{x\infty} h(x)
\sum_{r=0}^{n}
 \frac{\alpha^r q^{2(r^2 -r)}}{[2r]!!} x^{2r}
 \ga
\ee
Using (\ref{aif1}) we thus get the integral in terms
of the coefficients of the power series of $h(x)$ as
\be
 I(\ga) \sum_{r=0}^{\infty} \sum_{s=0}^{\infty}
\frac{[2r + 2s -1]!!}{[2r]!!}
\alpha^{-s} q^{-2(s^2+2rs +r)} h_{2s}
\ee
where $h(x) = \sum_{s=0}^{\infty} h_s x^s$. Note that the two
summations do not in general commute. The global integral
of a function $h(x) \in \fun $ can thus be written
\be
\int_{-x\infty}^{x\infty}  h(x)
 = I(\ga) S_{\alpha}.h\vert_{\small{x=0}}
\ee
where $S_{\alpha} \in \cal{Q}$ is the operator
\be
S_{\alpha} =
 \sum_{r=0}^{\infty} \sum_{s=0}^{\infty}
\frac{[2r + 2s -1]!!}{[2r]!![2s]!}
\alpha^{-s} q^{-2(s^2+2rs +r)} {\pax}^{2s}.
\ee
For sufficiently well behaved power series in $\fun $, which we shall call
integrable functions,
the above global integral
is a finite multiple of $I(\ga)$ and, using (\ref{gig}), we have also
\be
\int_{-x\infty}^{\infty} \pax h(x) = 0.
\ee
Recall that instead of the Gau{\ss}ian one could also use
another rapidly decreasing function $k$. We would then arrive at a
global integration formula that expresses integrals as multiples
of the global integral $I(k)$.

\section{The $n$ dimensional case}

We will now generalise some of our algebraic integration
techniques for the $n$ dimensional case.

To begin with we show for the $SL_q(n)$ case
how operators which act as indefinite integration
can again be found in the algebra $\cal{Q}$ generated by
the differentiation and multiplication operators $\partial^i$
and ${\bf x}_j$, $(i,j= 1,2,...,n)$. We again consider
functions of rapid decrease and integration limits that are scaled to infinity.

Then we develop the general case of Gau{\ss}ian
induced integration for quantum planes associated to R-matrices. Unlike the
one-dimensional case, we
do not discuss the integrability and rapid decay
properties of the Gau{\ss}ians in detail,  taking instead a more algebraic
line.
One still has precise integration formulae without knowing details of the
Gau{\ss}ian itself or its
integral. We give the examples of $q$-Euclidean space, $q$-Minkowski space and
$SL_q(n)$ quantum planes explicitly.
A representation theoretic approach will be followed elsewhere, including also
the problem of expressing the $n$-dimensional integral as a trace along the
lines of (\ref{trace}).

\subsection{Integration operators on $SL_q(n)$ quantum planes}
The commutation relations of the $SL_q(n)$ - comodule algebra
 $\cal{Q}$ of multiplication and differentiation operators
 read \cite{WesZum:cov}:
\begin{eqnarray}
{\partial}^i {\partial}^j - q {\partial}^j
{\partial}^i & =  & 0 \mbox{ \quad for
\quad } i < j \label{eins}\\
{\bf x}_i {\bf x}_j -
q {\bf x}_j {\bf x}_i & =  & 0
\mbox{ \quad for \quad } i > j \label{zwei} \\
{\partial}^i {\bf x}_j  - q {\bf x}_j {\partial}^i & =  & 0
\mbox{ \quad for \quad } i \ne j \label{drei} \\
{\partial}^i {\bf x}_i
- q^2 {\bf x}_i {\partial}^i  & =  & 1
+ (q^2 - 1) \sum\limits_{j<i} {\bf x}_j {\partial}^j
\label{vier}
\end{eqnarray}
These relations and their complex conjugates
appeared for operators on $q$-deformed Bargmann Fock space in
\cite{Kem:sym}\cite{Kem:int}.

We see that the first dimension can be identified with
the one dimensional case that we have considered so far.

The algebra is again naturally represented on the function algebra
${\fun}$ of power series in the generators $x_i$, obeying
the same commutation relations as the multiplication operators
${\bf x}_i$ i.e. (\ref{zwei}). We want to identify
 $\int_0^{x_i}$
or ${\partial^i}^{-1}$ as an element
of the algebra $\cal{Q}$. Writing formally
\be
\int_0^{x_i} = {\partial}_i^{-1} =
 {\bf x}_i \frac{1-q^2}{1- (1-(1-q^2)\partial^i {\bf x}_i)}
\ee
we note that the last term is actually well defined as a power series.
This is because the action of the operator
$ (1-(1-q^2)\partial^i {\bf x}_i)$ is
\be
(1-(1-q^2)\partial^i {\bf x}_i).x_1^{r1}\cdot \cdot \cdot x_i^{r_i}
\cdot \cdot \cdot x_n^{r_n} =
(1-q^{2(r_1 +...+r_{i-1})}(1-q^{2(r_i+1)}))
x_1^{r1}\cdot \cdot \cdot x_i^{r_i}
\cdot \cdot \cdot x_n^{r_n}.
\label{fs}
\ee
The operator is thus diagonal in this basis of $\fun$, and
we read off that its eigenvalues are all of absolute value smaller
than $1$. This allows its expansion as a geometrical series:
\be
\int_0^{x_i} = (1-q^2) {\bf x}_i \sum_{n=0}^{\infty}
 (1-(1-q^2)\partial^i {\bf x}_i)^n
\ee
We thus have the desired properties:
\be
\partial^i \int_0^{x_i} f = f \qquad \quad \forall f \in \fun \label{rinv}
\ee
\be
\int_0^{x_i} \partial^i x_i f = x_i f \qquad \quad \forall f \in \fun.
\label{linv}
\ee
We used the factor $x_i$ in order to prevent $\int_0^{x_i}$ from
acting on $0$ which could otherwise occur through
 the annihilation of $f$ by the action of $\pa{i}$. It is
not very difficult to prove these properties
 also directly by using the above basis of ordered polynomials. Actually we see
that
due to this choice of basis, those derivatives that arise from
the term on the rhs of (\ref{vier}) do not find co-ordinates to
act on. This means that for ordered polynomials in the evaluation
of
\be
\int_0^{x_1} \int_0^{x_2} ... \int_0^{x_n}
x_1^{r_1}\cdot \cdot \cdot x_n^{r_n}
\ee
one can neglect this term on the rhs of (\ref{vier}) so
that we have in every dimension the effective
commutation relation
\be
{\partial}^i {\bf x}_i
- q^2 {\bf x}_i {\partial}^i = 1
\label{ef1dim}
\ee
which is of the same form as in the one dimensional relation.
It is thus a special feature of these ordered polynomials
that the operators $\int_0^{x_i}$ which act on them are
built of operators
which  effectively
obey the one dimensional commutation relation.
Even then, the  $\int_0^{x_i}$
 are  noncommuting with each other and with the co-ordinate functions. We will
 find a similar phenomenon of effective decoupling
of the integrations on ordered polynomials in the $SL_q(n)$ case of the
$n$-dimensional Gau{\ss}ian induced integration method.

Since the operator $\int_0^{x_i}$ is the inverse of the operator $\partial^i$
in the representation of the algebra $\cal{Q}$ as operators on the function
algebra $\fun$,
we immediately get the commutation relations:
\begin{eqnarray}
{\partial}^j \int_0^{x_i} - q \int_0^{x_i} {\partial}^j
& =  & 0 \mbox{ \quad for
\quad } i < j \\
q{\partial}^j \int_0^{x_i} -  \int_0^{x_i} {\partial}^j
& =  & 0 \mbox{ \quad for
\quad } i > j \\
\int_0^{x_i} \int_0^{x_j} -
q \int_0^{x_j} \int_0^{x_i} & =  & 0
\mbox{ \quad for \quad } i > j \label{qintint} \\
 {\bf x}_j \int_0^{x_i}  - q \int_0^{x_i} {\bf x}_j & =  & 0
\mbox{ \quad for \quad } i \ne j  \\
 {\bf x}_i \int_0^{x_i}
- q^2 \int_0^{x_i}  {\bf x}_i  & =  &  \int_0^{x_i} \int_0^{x_i}
+ (q^2 - 1) \sum\limits_{j<i}{\bf x}_j
\pa{j} \int_0^{x_i}\int_0^{x_i}
\label{ibp}
\end{eqnarray}
which could of course also be written in R-matrix notation.
Note that (\ref{ibp}) describes integration by parts.
$\int_0^{x_i}$ behaves like  $x_i$.

In complete analogy with the one dimensional case one can
also use scaling operators $L_i$ to define
functions that vanish at infinity, rapidly decreasing functions
and the scaling of integration limits to infinity. We arrive then at
operators $\intinf {x_i}$ and could define for example
\eqn{intprod}{ \int=\intinf {x_1}...\intinf {x_n}.}
for a global integration operator over the whole space. Using the relations
(\ref{qintint}) we can always bring any of the $\intinf {x_i}$  to the right,
so translation invariance in the form $\int\del^i =0$ is assured. We note that
integration on $SL_q(n)$ quantum planes as an iteration of 1-dimensional
Jacskon integrals has been proposed previously in \cite{KorVak:spa} from a
different point view.

This is one very concrete approach to constructing an integral given here in
the $SL_q(n)$ case. Next we develop next a more powerful Gau{\ss}ian approach,
but will see in Section~5.3.3 that it is consistent with the above concrete
one.

\subsection{Preliminaries on braided differentiation and integration}

Here we collect some formulae of general
braided differential calculus \cite{Ma:fre} which we will need in the next
section, where we develop
integration for this setting. The data we need are
matrices $R,R^{\prime} \in M_n \otimes M_n$ that fulfill:
\be
R_{12} R_{13} R_{23} = R_{23} R_{13} R_{12}\label{QYBE}
\ee
\be
R_{12} R_{13} R_{23}^{\prime} = R_{23}^{\prime} R_{13} R_{12}, \qquad
R_{12}^{\prime} R_{13} R_{23} = R_{23} R_{13} R_{12}^{\prime}\label{RRR'}
\ee
\be
R_{21} R^{\prime}_{12} = R^{\prime}_{21} R_{12}, \qquad\label{RR'}
(PR + 1)(PR^{\prime}-1)=0
\ee
where $P$ is the permutation matrix. Examples of matrices $R^{\prime}$
can be constructed from the minimal polynomial of $PR$. We
further assume that $R$ be invertible in two ways:
\be
\exists R^{-1},\tilde{R}: \qquad {R^{-1}}^i{}_a{}^j{}_b R^a{}_k{}^b{}_l =
\delta^i{}_k
\delta^j{}_l={\tilde{R}}^a{}_k{}^j{}_b R^i{}_a{}^b{}_l. \label{Rinv}
\ee
The braided differential calculus then implies the following
commutation relations in the algebra $\cal{Q}$ of
multiplication and differentiation operators:
\be
x_i x_j - x_b x_a {R^{\prime}}^a{}_i{}^b{}_j =0
\label{polys}
\ee
\be
\pa{i}\pa{j} - {R^{\prime}}^i{}_a{}^j{}_b\pa{b}\pa{a} = 0
\label{dd}
\ee
and
\be
\pa{i} x_j - x_a R^a{}_j{}^i{}_b \pa{b} = \delta^i{}_j\label{Rleibniz}
\ee
or equivalently:
\be
x_i \pa{j} - {\tilde{R}}^b{}_i{}^j{}_a\pa{a}x_b =
- {\tilde{R}}^a{}_i{}^j{}_a.
\label{dvr}
\ee
Our convention here and below is that repeated indices $a,b,c$ etc., are to be
summed, but indices $i,j,k$ etc, remain free. Again, the algebra
(\ref{polys})--(\ref{dvr}) acts naturally as operators on the
function algebra generated by generators $x_1,...,x_n$,
obeying the position co-ordinate relations (\ref{polys}). We no longer use
bold-face to denote $x_i$ as operators by multiplication.

The goal is to give a method for translation-invariant integration in this
general R-matrix setting. Let us note that the question of existence of such a
map $\int$ is not really an issue, at least in the global case. The position
co-ordinates $x_i$ form a braided-Hopf algebra\cite{Ma:poi} and just as for
usual groups and quantum groups, there are quite general ways to argue
existence of $\int$. The problem is rather one of explicit evaluation, which is
what we provide in the next section.

Before proceeding to this Gau{\ss}ian integration, we note that one could also
try to follow the concrete approach of Section~5.1. The key idea is to find
partial inverses $\int_0^{x_i}$ for the differentiation operators $\pa{i}$ in
the spirit of (\ref{rinv})--(\ref{linv}). They may no longer be expressible in
the associated Heisenberg algebra generated by $\del^i,x_j$ but (\ref{dd})
still yields their commutation relations
\be
\int_0^{x_j} \int_0^{x_i} - {R^{\prime}}^i{}_a{}^j{}_b
\int_0^{x_a} \int_0^{x_b} = 0.
\label{intintrel}
\ee
Next one has to scale these to infinity. Since the relations (\ref{polys}) are
quadratic, one possibility is $L^{-\infty}$ with  $L(x_i)=q^2 x_i$  for a
parameter $q< 1$. Other scaling methods are also possible. Finally, one can
write down products such as (\ref{intprod}). If the scaled form of the
relations (\ref{intintrel}) are sufficiently nice that one can move any of the
integrations to the right, one would still have $\int\del^i=0$ as required. In
general however, the required form may be more complicated and has to be
analysed case by case. By contrast, the following Gau{\ss}ian method yields
quite general formulae without knowing details of $R$ and $R'$.

\subsection{Gau{\ss}ian-induced integration for general R-matrices}

We have already analysed the Gau{\ss}ian method in some detail in the
one-dimensional case and seen that a Gau{\ss}ian exists in this case, with the
right properties of rapid decay etc in a suitable representation. Our strategy
in this section is to assume a similar Gau{\ss}ian now in $n$-dimensions. We
will not try to give it explicitly or prove its decay properties. These  are
topics for further work. Remarkably, the Gau{\ss}ian approach yields numbers
that do not depend on these details as long as we are content to integrate only
functions against this Gau{\ss}ian weight.

We let $\eta_{ij} \in M_n$ be a matrix and assume that a solution $g_\eta$
of the equation
\be
-\eta_{ia}\pa{a} g_\eta = x_i. g_\eta
\label{heateq}
\ee
exists and is (at least formally) rapidly decreasing and integrable with
respect to an integration $\int$, to be determined. We will not try to
determine
\[ I(g_\eta)=\int g_\eta\]
itself but only integrals of $fg_\eta$ where $f$ is a polynomial in our
non-commutative position co-ordinates.  We assume furthermore that
\be
\int \pa{i} f g_\eta = 0
\qquad \mbox{ for all polynomials} f \in \mbox{$\fun$},\quad\forall i=1,...n
\label{intdfg}
\ee
which says that we neglect boundary terms of the form (polynomial $\cdot$
$g_\eta$).

We show now that these assumptions uniquely determine $\int fg_\eta$ in terms
of $\eta$ and our R-matrix data. Indeed, we show that
this takes the form
\be
 \int fg_\eta = \Zint[f] I(g_\eta)
\label{eval}
\ee
where $\Zint[f] \in \C$ for any polynomial $f\in \fun$.

We prove the existence of the linear functional
$\Zint$ by giving an explicit way to calculate it.
Consider the global integral with an
arbitrary multinomial $f=x_{i_1}\cdots x_{i_m}$ say. The indices here need not
be distinct. Then
\align{\int x_{i_1}\cdots x_{i_m} g_\eta\equad &&=-\int x_{i_1}\cdots
x_{i_{m-1}}\eta_{i_m a}\del^a g_\eta\\
&&=-\int x_{i_1}\cdots x_{i_{m-2}}\eta_{i_m
a}\left(-\tilde{R}^b{}_{i_{m-1}}{}^a{}_b+\tilde{R}^c{}_{i_{m-1}}{}^a{}_d \del^d
x_c\right)g_\eta}
using (\ref{heateq}) and (\ref{dvr}). We think of $x_i$ as operators acting to
the right by multiplication. The first term here is the integral of a monomial
of degree $m-2$ times $g_\eta$. We can use (\ref{dvr}) in the same way again
for $x_{i_{m-2}}\del^d$ which generates another integral of a monomial of
degree $m-2$ times $g_\eta$, and moves $\del$ further to the left. We repeat
this until the term containing $\del$ is of the form $\int\del \cdots g_\eta$,
which then vanishes by (\ref{intdfg}). Hence $\int x_{i_1}\cdots
x_{i_{m}}g_\eta$ is expressed explicitly in terms of the integration of
monomials of degree two less, times $g_\eta$. Since $\int g_\eta$ is of the
required form with $\Zint[1]=1$, we see by induction that $\int x_{i_1}\cdots
x_{i_m} g_\eta=\Zint[x_{i_1}\cdots x_{i_m}]I(g_\eta)$ where
$\Zint[x_{i_1}\cdots x_{i_m}]$ is a complex number built from the matrices
$\eta$ and $\tilde{R}$. This is our Gau{\ss}ian integration method. Note that
we did not require anywhere that the algebra of co-ordinates was commutative.

Clearly, this method works whenever the differentials and co-ordinates enjoy a
relation of the general form (\ref{dvr}). Moreover, our derivation is
independent of the detailed form of both $g_\eta$ and $\int$. Any choice of
these such that (\ref{heateq}) and (\ref{intdfg}) hold will yield the same
values $\Zint[f]$. This $\Zint[f]$ is the {\em Gau{\ss}ian-weighted} integral
of a polynomial in our co-ordinate algebra. Of course, details of $g_\eta$
would be needed if we wanted to know exactly what functions $fg_\eta$ we have
integrated with respect to $\int$. Classically, this class would be dense in
the set of Riemann-integrable functions if one takes the usual Euclidean
metric. One would also like to know exactly how the class of integrable
functions with respect to $g_\eta$ is related to the class with respect to
$g_{\eta'}$ for a different choice of metric. One can expect that these various
questions have algebraic answers the along lines developed in Section~4 for the
one-dimensional case. These finer points of `quantum analysis' will be
investigated elsewhere.

We specialise now to an important class of metrics $\eta$, namely ones that are
quantum group covariant. The co-ordinate algebra generated by the $x_i$, the
algebra of differentials $\del^i$ etc, were introduced in $\cite{Ma:poi}$ in a
manifestly covariant way under the usual matrix quantum group
$\vect=\{t^i{}_j\}$ associated to $R$. We suppose here that $R$ is of the
regular type so that there is an antipode, i.e. an inverse quantum matrix
$\vect^{-1}$. The co-ordinates transform as covectors by $x_j\to x_it^i{}_j$
and the differentials transform as vectors using $\vect^{-1}$. In Section~6.3
we will include a dilaton $\dila$ also in order to induce the correct braiding
there. In order to remain in this covariant setting, we assume now that $\eta$
is invariant in the sense
\eqn{etainv}{ \eta_{ab}t^a{}_i t^b{}_j=\eta_{ij}.}
Using the properties of the dual-quasitriangular structure induced by $\lambda
R$, \cite{Ma:lin}, one finds easily the useful identities
\eqn{etaR}{\eta_{ja}R^l{}_i{}^a{}_k=\lambda^{-2
}R^{-1}{}^l{}_i{}^a{}_j\eta_{ak},\quad
\eta_{ja}\tilde{R}^l{}_i{}^a{}_k=\lambda^2 R^l{}_i{}^a{}_j\eta_{ak}.}
\begin{equation}
\eta_{ia}\eta_{jb}R^a{}_m{}^b{}_n=R^a{}_i{}^b{}_j
\eta_{am}\eta_{bn}, \quad
\eta_{ai}R^a{}_j{}^l{}_k=\lambda^{-2}R^{-1}{}^a{}_i{}^l{}_k\eta_{ja}
\end{equation}
Here $\lambda$ is a constant that converts $R$ in (\ref{RR'}) into the quantum
group normalisation as explained in \cite{Ma:poi}. Using
(\ref{etaR}), we write (\ref{Rleibniz}) as
\eqn{Rleibnew}{ \del_i x_j=\eta_{ij}+ x_a \del_c \lambda^{-2}
R^{-1}{}^a{}_j{}^c{}_i,\quad {\rm i.e.}\quad
\del_1\vecx_2=\eta_{12}+\vecx_1\del_2 \lambda^{-2} (PR)^{-1}_{12}}
where the right hand form is in a standard compact notation in which the
numerical suffices stand for the positions of repeated tensor indices. We use
bold-face here to denote an entire vector or matrix of generators.
The inverse relation (\ref{dvr}) is then clearly
\eqn{dvrnew}{  \vecx_1\del_2=(-\eta_{12}+\del_1\vecx_2)(PR)_{12}\lambda^2.}

Finally, we check a kind of `quantum integrability' condition for our heat
equation
$\del_i g=-x_i g$. Recall that the usual integrability condition for a partial
differential equation comes from requiring commutativity of partial
derivatives. In our case we require (\ref{dd}) and find the condition for this
by computing
\align{&&\equad \del_1\del_2 g= -\del_1\vecx_2
g=-\eta_{12}g-\vecx_1\del_2\lambda^{-2}(PR)^{-1}_{12}g\\
&&=-\eta_{12}g+\vecx_1\vecx_2\lambda^{-2}(PR)^{-1}_{12}g
=-\eta_{12}g+\vecx_1\vecx_2\lambda^{-2}(PR')_{12}(PR)^{-1}_{12}g\\
&&=-\eta_{12}g+\vecx_1\vecx_2\lambda^{-2}(PR)^{-1}_{12}(PR')_{12}g
=\eta_{12}((PR')_{12}-1)g+\del_1\del_2 (PR')_{12}g}
using (\ref{Rleibnew}), the first of (\ref{RR'}), and (\ref{polys}). For the
quadratic term in derivatives to vanish, we see that we require  $\del_i$ to
obey the algebra (\ref{polys}) like the $x_i$. Since $\del^i$ obey (\ref{dd}),
the conditions for `quantum integrability' of our quantum heat equation become
\eqn{etaR'}{\eta_{ia}\eta_{jb}R'{}^a{}_m{}^b{}_n=R'{}^a{}_i{}^b{}_j
\eta_{am}\eta_{bn},\quad \eta_{ab}R'{}^b{}_i{}^a{}_j=\eta_{ij}.}
These are  natural conditions on the metric and hold in important examples such
as $q$-Minkowsksi space\cite{Mey:new}, where (\ref{etaR'}) were introduced in
connection with the isomorphism between spacetime vectors and covectors.

We therefore require our invariant metric $\eta$ to obey (\ref{etaR'}) also. In
this case we can reasonably expect that a Gau{\ss}ian $g_\eta$ can be found at
least for sufficiently well-behaved $R$. We have already given general
arguments for the
integral $\int$. Putting these assumptions together we have:

\begin{theorem} Let $R,R'$ and an invariant matrix $\eta$ obey the conditions
above. Then
\[ \Zint[\vecx_1\cdots\vecx_m]=\left(\int \vecx_1\cdots\vecx_m g_\eta \right)
\left(\int g_\eta\right)^{-1}\]
is a well-defined linear functional from the algebra (\ref{polys}) generated by
the co-ordinates $\{x_i\}$ to $\C$ and can be computed inductively by
\[\Zint[1]=1,\quad  \Zint[\vecx]=0, \quad
\Zint[\vecx_1\vecx_2]=\eta_{12}(PR)_{12}\lambda^2\]
\[\Zint[\vecx_1\cdots \vecx_m]=\sum_{i=0}^{m-2} \Zint[\vecx_1\cdots \hat
{\vecx_{i+1}}\hat{\vecx_{i+2} }\cdots\vecx_{m}]\Zint[\vecx_{i+1}\vecx_{i+2}]
(PR)_{i+2\ i+3}\cdots (PR)_{m-1\ m}\lambda^{2(m-2-i)}\]
where $\hat{\ }$ denotes omission. We call $\Zint$  the {\em
Gau{\ss}ian-weighted integral functional} on the braided space. It is invariant
under the background quantum group.
\end{theorem}
\proof The formulae here come from the Gau{\ss}ian integration method as
explained above. We have
\align{\int \vecx_1\cdots\vecx_m
g_\eta\equad&&=-\int\vecx_1\cdots\vecx_{m-1}\del_m g_\eta\\
&&=-\int\vecx_1\cdots\vecx_{m-2}\left(-\eta_{m-1\
m}+\del_{m-1}\vecx_m\right)(PR)_{m-1\ m}\lambda^2 g_\eta}
much as before, but this time using (\ref{dvrnew}) and the compact notation. We
repeat this again for the $\vecx_{m-2}\del_{m-1}$ etc., until we reach
$\int\del=0$. For $m=2$ we have
$\Zint[\vecx_1\vecx_2]$ in terms of $\eta$, which we then use to give the
general form stated. The invariance corresponds to the assumption that $\int,
g_\eta$ should be invariant under the quantum group. \endproof

This gives a way to compute the integral knowing only the R-matrix data and
$\eta$. In principle, one can in fact take Theorem~5.1 as an inductive
definition of $\Zint$, verifying directly from the initial data
(\ref{QYBE})--(\ref{RR'}) and (\ref{etainv}), (\ref{etaR'}) that it is
well-defined and invariant. Thus, to see invariance, we use the FRT relations
for the quantum group \cite{FRT:lie}cf\cite{Dri} in the form $PR\vect_1\vect_2
=\vect_1\vect_2 PR$. Then
\[\Zint[\vecx_1\vecx_2\vect_1\vect_2]=\lambda^2\eta_{12}(PR)_{12}\vect_1\vect_2
=\lambda^2\eta_{12}\vect_1\vect_2(PR)_{12}=\Zint[\vecx_1\vecx_2]\]
using (\ref{etainv}) for the last equality. We then prove the general case in
the same way by induction using the stated formula for
$\Zint[\vecx_1\cdots\vecx_m]$ in terms of lower ones and $PR$. We can likewise
verify directly that $\Zint[\vecx_1\vecx_2]$ is well-defined as
\[ \Zint[\vecx_1\vecx_2PR']=\lambda^2\eta_{12} PR PR'= \lambda^2
\eta_{12}PR'PR=\Zint[\vecx_1\vecx_2] \]
using  (\ref{RR'}) and (\ref{etaR'}). For the higher orders the direct proof is
rather involved and will not be attempted here. In lieu of a formal proof, we
have given general arguments based on integrability of (\ref{heateq}) to expect
this for any reasonable $R,R',\eta$. Since the construction of $\Zint$ is
purely algebraic, any further subtleties about convergence of powerseries and
`quantum analysis' which would be needed for $\int$ and $g_\eta$ themselves, do
not apply. Moreover, it is easy enough to compute $\Zint$ by our formula and
check in each example that it is well-defined.

We remark also that the factorisation in Theorem~5.1 (and a similar one
involving $\tilde R$ in the general non-invariant case) is remeniscient of the
well-known factorisation property of the $S$-matrix in certain exactly solvable
quantum statistical systems. It could be viewed as some kind of quantum or
braided correlation function and $\Zint[\vecx_1\vecx_2]$ as the 2-point
function. On the other hand, the $x_i$ are for us are non-commutative position
co-ordinates, such as those of $q$-Euclidean and  $q$-Minkowski space.

\subsubsection{Example of $q$-Euclidean space}

For $q$-Euclidean space, we use the definition in \cite{Ma:euc} as twisting
$\bar M_q(2)$ of the usual $2\times 2$ quantum matrices. We have generators
$a,b,c,d$ and relations
\ceqn{qeucl}{ba=qab,\quad ca=q^{-1}ac,\quad da=ad,\quad db=q^{-1}bd\quad
dc=qcd\\
bc=cb+(q-q^{-1})ad.}
This is actually isomorphic to the usual $M_q(2)$ by a permutation of the
generators, so one can
regard the following as integration on this with its additive structure as
introduced in \cite{Ma:add}.

It is easy to find the quantum metric from the data in \cite{Ma:euc}, to which
we refer the reader. The $R,R'$ are built from two copies of the usual
$SL_q(2)$ R-matrix. The metric is built from the invariant metric associated to
each of these and comes out as
\eqn{etaeucl}{\eta_{ij}=\pmatrix{0&0&0&1\cr 0&0&-q^{-1}&0\cr 0&-q&0&0\cr
1&0&0&0}}
It is easy to see that it obeys the conditions (\ref{etaR'}) needed for
integrability. We need $\lambda=q^{-1}$ to connect with the quantum group
normalisation.

{}Using Theorem~5.1, we find that the `two-point function' comes out as
proportional to the metric,
\eqn{Zeta}{ \Zint[x_ix_j]\equiv{\int x_ix_j g\over\int g}= q^{-4}\eta_{ij}}
while the next lowest order is therefore
\eqn{ZZeta}{\Zint[x_ix_jx_k x_l]\equiv {\int x_{i_1}x_{i_2}x_{i_3}x_{i_4}
g\over\int g}= q^{-8}\eta_{ij}\eta_{kl}+q^{-10}\eta_{ia}\eta_{jb}
R^a{}_{k}{}^b{}_{l}+q^{-12}\eta_{ia}\eta_{bc}R^b{}_{j}{}^a{}_d
R^c{}_{k}{}^d{}_{l}}
where $R={\bf R}_+$ in \cite{Ma:euc}. Similarly for higher orders, each
involving more powers of $\eta$. It is a non-trivial fact that these are indeed
well-defined linear maps $\bar M_q(2)\to \C$. This can easily
be checked for lower orders by computer calculations. We have done so using the
computer package REDUCE. The non-zero integrals
of quartic expressions come out as
\cmath{ \Zint[abcd]=-q^{-1},\quad \Zint[acbd]=-q^3,\quad \Zint[a^2d^2]=q^2+1\\
 q^2\Zint[b^2c^2]=q^{-2}\Zint[c^2b^2]=\Zint[bc^2b]=\Zint[cb^2c]=q^{2}+1,\quad
q^2\Zint[bcbc]=\Zint[cbcb]=q^4+1,}
times an overall factor $q^{-10}$ and plus the others needed for consistency
under the  first five relations in (\ref{qeucl}). One can see that the
remaining relation in (\ref{qeucl}) is respected.

We can also consider the spacetime and radius co-ordinates
\[ t={a-d\over \imath},\quad x={c-qb}, \quad y={c+qb\over \imath},\quad
z={a+d},\quad r^2=ad-q^{-1}bc\]
in terms of the matrix generators $a,b,c,d$. Then
\[ \Zint[t^2]=\Zint[z^2]={2\over q^4},\quad
\Zint[x^2]=\Zint[y^2]=q^2\Zint[r^2]={[2]\over q^4}\]
\[\h\Zint[t^4]=\h\Zint[z^4]={3[2]\over q^{10}},\quad
\h\Zint[x^4]=\h\Zint[y^4]=q^4\Zint[r^4]={[3]!\over q^{10}}\]
where $[n]\equiv {q^{2n}-1\over q^2-1}$ is the usual $q$-integer. These results
agree with the classical values after setting $q=1$ where $r^2={1\over
4}(t^2+x^2+y^2+z^2)$. We see that the Gau{\ss}ian-weighted integral $\Zint$ has
a
degree of spherical symmetry even in the non-commutative case.

We remark that in this example (and other cases such as the $q$-Minkowski space
in the next section) one has the additional $R$-matrix relations
\be
R^i{}_a{}^j{}_b \eta^{ab} = q^{-2} \eta^{ji}
\label{einsa}
\ee
\begin{equation}
\eta_{ia} R^{\prime}{}^j{}_k{}^a{}_l =
R^{\prime}{}^j{}_k{}^a{}_i \eta_{aj}
\label{zweia}
\ee
with $\eta^{ab}$ being the inverse to $\eta_{ij}$. In this case one can compute
the
Gau{\ss}ian explicitly as
\be
g= \sum_{r=0}^{\infty} {1\over [r]!}\left({x\cdot x\over 1+q^{-2}}\right)^r
= e_{q^2}^{\frac{x\cdot x}{1+q^{-2}}}
\label{dreia}
\ee
where $[r]=(1-q^{2r})/(1-q^2)$ and $x\cdot x=x_ax_b\eta^{ab}$.
We use (\ref{einsa}) to compute the commutation relations
\be
\partial_j(x\cdot x)=(1+q^{-2})x_j + q^2 x\cdot x \partial_j
\ee
and (\ref{zweia}) to show that $x\cdot x$ commutes with $x_j$, after which
(\ref{dreia}) follows. This makes contact with the specific Gau{\ss}ian for
$SO_q(n)$-quantum planes in \cite{Fio:sym}. Our $R$-matrix formulation is
however, more general.

\subsubsection{Example of $q$-Minkowski space}

For $q$-Minkowski space, we use the definition as the algebra $BM_q(2)$ of
$2\times 2$ braided hermitian matrices\cite{Ma:exa} with generators $a,b,c,d$
and relations
\ceqn{qmink}{ba=q^2ab,\quad ca=q^{-2}ac,\quad da=ad,\qquad
bc=cb+(1-q^{-2})a(d-a)\\
db=bd+(1-q^{-2})ab,\quad cd=dc+(1-q^{-2})ca}
The $R,R'$ description and the necessary metric are also known. We use the
recent formulation of \cite{Mey:new}, but see also
\cite{CWSSW:lor}\cite{OSWZ:def}. The quantum metric we need is
\eqn{etamink}{\eta_{ij}=\pmatrix{0&0&0&1\cr 0&0&-q^{-2}&0\cr 0&-1&0&0\cr
1&0&0&1-q^{-2}}}
and is shown in \cite{Mey:new} to obey the key equations (\ref{etaR'}) for
integrability. Here $\lambda=q^{-1}$.

{}From Theorem~5.1 the `two-point function' again comes out as proportional to
the metric and with the same factors as in the
Euclidean case, i.e. we have (\ref{Zeta})--(\ref{ZZeta}) again. The metric is
now (\ref{etamink}) and  $R=R_L$ in \cite{Mey:new}. Similarly for higher orders
in $\eta$. Again, it is a non-trivial fact that these are indeed well-defined
linear maps $BM_q(2)\to\C$. This has been verified to low orders using REDUCE.
For the  the non-zero integrals
of quartic expressions, we have
\cmath{ \Zint[abcd]=\Zint[abdc]=\Zint[adbc]=\Zint[bcad]=-q^{-2},\quad
\Zint[acbd]=\Zint[acdb]=\Zint[adcb]=-q^2 \\
 q^2\Zint[bdcd]=q^2\Zint[dbdc]=\Zint[cdbd]=\Zint[dcdb]=1-q^2\\
q^4\Zint[bcd^2]=q^4\Zint[dbcd]=q^4\Zint[d^2bc]=q^2\Zint[cd^2b]={1-q^4}\\
q^2\Zint[bd^2c]=\Zint[cbd^2]=\Zint[dcbd]=\Zint[d^2cb]=-\h
\Zint[d^4]=-(q-q^{-1})^2\\
\Zint[ad^3]=(1+2q^2)(1-q^{-2}),\quad  q^4\Zint[bcbc]=q^2\Zint[cbcb]={q^4+1}\\
\Zint[a^2d^2]=q^4\Zint[b^2c^2]=\Zint[c^2b^2]=q^2\Zint[bc^2b]
=q^2\Zint[cb^2c]={q^2+1}}
times an overall factor $q^{-10}$ and plus the others cases needed for
consistency under the  first three relations in (\ref{qmink}). One can see that
the remaining three relations in (\ref{qmink}) are respected.

We can also consider the spacetime and radius co-ordinates
\[ t=q^{-1}a+qd,\quad x=b+c,\quad y={b-c\over \imath},\quad z={d-a},\quad
r^2=ad-q^2cb\]
in terms of the matrix generators $a,b,c,d$. Then
\[ \Zint[t^2]=\Zint[r^2]={[2]\over q^4},\quad
\Zint[x^2]=\Zint[y^2]=\Zint[z^2]=-{[2]\over q^6}\]
\[\h\Zint[z^4]={3[2]\over q^{12}},\quad \h\Zint[t^4]=\h q^4\Zint[x^4]=\h q^4
\Zint[y^4]=\Zint[r^4]={[3]!\over q^{10}}\]
which is consistent at $q=1$ with $r^2={1\over 4}(t^2-x^2-y^2-z^2)$. We see
that the Gau{\ss}ian-weighted integral $\Zint$ is quite similar to the
Euclidean
one in its values, except for the sign in the spacelike directions.

This $q$-Minkowski space  is a striking example of our algebraic machinery
because classically of course, the Gau{\ss}ian has far from rapid decay in the
space-like directions! The integrals $\int x_ix_jx_kx_l g$ etc, and $\int g$
are both infinite but
their ratio $Z[f]={\int fg\over\int g}$ is perfectly well-defined by the
algebraic formulae in Theorem~5.1. This is familiar in the context of path
integrals in quantum theory, but applies just as well for Gau{\ss}ian-weighted
integrals in spacetime. Moreover, it applies just as well in the $q$-deformed
non-commutative setting. From an algebraic point of view, the stucture of
$q$-Minkowski space is in fact strictly related to the $q$-Euclidean structure
by a `quantum Wick rotation'\cite{Ma:euc}.

\subsubsection{Example of $SL_q(n)$ quantum planes}

Here we demonstrate the Gau{\ss}ian method for the
$SL_q(n)$ quantum plane, using the usual Euclidean metric
$\eta_{ij}=\alpha^{-1}\delta_{ij}$. One can check that even though this is not
invariant, the special form of the relations (\ref{eins})--(\ref{zwei}) is such
that (\ref{heateq}) is consistent with the relations of the algebras, i.e. that
the equation is `quantum integrable'. As before, we do not actually compute the
Gau{\ss}ian $g$.

We compute $\Zint[f]$ where $f$ is a polynomial in our non-commuting
co-ordinates $x_i$. Without loss of generality we can assume that its terms are
conveniently ordered in decreasing co-ordinate number. Then the Gau{\ss}ian
method gives on each such term:
\align{
\Zint[f] I(g) \equad&&=
\int x_n^{r_n}  \cdots x_2^{r_2} x_1^{r_1} g\\
&&= -\alpha^{-1} \int x_n^{r_n}  \cdots x_2^{r_2} x_1^{r_1-1} \pa{1} g\\
&&=  \alpha^{-1} q^{-2r_1} [r_1-1]
\int x_n^{r_n}  \cdots x_2^{r_2} x_1^{r_1-2} g\\
&&=
\alpha^{-r_1/2} q^{-2(r_1 +...+ 4+2))} [r_1-1]!!
\int x_n^{r_n}  \cdots x_2^{r_2} x_1^{r_1-2} g}
for $r_1$ even, and $0$ for $r_1$ odd. Inductively then, we have
\be
\Zint[x_n^{r_n}  \cdots x_2^{r_2} x_1^{r_1}]=  \alpha^{-(r_1+...+r_n)/2}
q^{-(r_1^2 +...+ r_n^2)/2}
[r_1-1]!!...[r_n-1]!! \label{Vsln}
\ee
The choice of ordering ensured that the derivatives produced through the rhs.
of (\ref{vier})
did not find co-ordinates to act on. In this way the problem
decomposed into $n$ integrations where each of them is performed
effectively like a $1$ dimensional integration.
Such an effective decomposition
 was of course to be expected since such a
phenomenon already appeared in the calculation that
had lead to the effective equation (\ref{ef1dim}).
Let us stress, however, that we ordered the polynomials only for
convenience. We would have come to exactly the same result if they
were not especially ordered.
 Our formalism, being adjusted to the noncommutative
setting, takes care of that automatically and works also in the
general case even where no ordering can be found that would decouple
the $n$-dimensional integration as it does here.

We recall also that we gave an explicit candidate for $\int$ in (\ref{intprod})
in this
$SL_q(n)$ case. Using it above would of course give the same answer since
$\Zint$ is independent of
the exact form of $\int$ as long as it is translation-invariant.  On the other
hand, we could also compute this particular
$\int$ another way. Namely, consider
\be
\int_{-x_n\infty}^{x_n\infty} f_n(x_n)g_n(x_n) \cdots
\int_{-x_1\infty}^{x_1\infty}
f_1(x_1) g_1(x_1)
\label{mu}
\ee
where the functions $g_1$,...,$g_n$ are Gau{\ss}ians, each dependent
only on one co-ordinate and fulfilling $\pa{i} g_i = -\alpha x_i g_i$.
This is actually the opposite ordering from the one that facilitated
the calculation in (\ref{fs}). With this choice, we can now integrate
using the Gau{\ss}ian method in each dimension separately as in Section~4.2.
This gives
\be
c\int_{-x_n\infty}^{x_n\infty} g_n(x_n) ...
\int_{-x_1\infty}^{x_1\infty} g_1(x_1)
\ee
Comparing Section~4.2 with (\ref{Vsln}) we see that $c=\Zint[f_n(x_n)\cdots
f_2(x_2)f_1(x_1)]$.
Thus we conclude that in the $SL_q(n)$ case of the global integral of
$f(x_1,...,x_n) g$ can also be obtained as the result
of $n$ independent Gau{\ss}ian integrations, namely by
writing $f$ as a sum of ordered terms and proceeding as in (\ref{mu}). This is
a kind of $q$-direct product theorem which says that we can compute our
Gau{\ss}ian-induced integrals as a product of lower-dimensional ones provided
suitable attention
is paid to ordering and $q$-factors. This can be expected to be a general
phenomenon for R-matrices of Hecke type which are the `gluing'\cite{MaMar:glu}
of lower-dimensional R-matrices.

\section{Fourier transform on braided spaces}

The  main work in this paper has been to develop a fairly general theory of
translation-invariant $q$-integration on braided vector spaces. The underlying
theory of braided-linear algebra, braided-coaddition, braided-differentiation
and the exponential map on these spaces has already been introduced (by the
second author\cite{Ma:lin}\cite{Ma:poi}\cite{Ma:fre}) and includes functions in
1 variable (the braided line), the quantum plane $yx=qxy$, $q$-Minkowski space,
etc. We are now in a position to combine our integration theory with this
existing theory and pick off an easy application, namely the theory of Fourier
transformation on such spaces.

We develop this first in an abstract diagrammatic language that works for any
braided group and translation-invariant `integration' functional on it. This
diagrammatic technique has been developed in \cite{Ma:introp} and several other
papers by the second author, and is by far the easiest way to prove our
results. After that, we examine how the theory looks in the 1-dimensional and
n-dimensional cases.

Note that the theory of quantum Fourier transform on quantum groups and certain
self-dual braided groups coming from quantum groups by transmutation was
already covered in \cite{LyuMa:bra}\cite{LyuMa:fou}. Our diagrammatic
definitions are compatible with this but formulated now without the need for
tensor categories with left and right duals and reconstruction theorems etc.
The present more general theory is needed to cover the braided vector spaces
such as quantum planes etc, which were not treated before. It should be
stressed that the abstract Fourier theory is quite straightforward and the
non-trivial part lies in the construction of the integration itself.

\subsection{Diagrammatic Fourier transform}

Let $B$ be a braided-Hopf algebra or `braided group' as introduced by the
second author\cite{Ma:exa}. We use the notation of \cite{Ma:lin} where the
product is written as $\cdot=\epsfbox{prodfrag.eps}$, the coproduct as
$\Delta=\epsfbox{deltafrag.eps}$ and the braiding as
$\Psi=\epsfbox{braid.eps}$. Other maps are written as boxes or labelled nodes
with the appropriate number of inputs and outputs. All maps are written as
flowing generally
downwards. The axioms of a braided-Hopf algebra are then recalled in Figure~1
(a).
Here $\eps$ is the counit and $\eta$ the unit. These have trivial braiding with
everything and hence can be written as going to/from nothing. In the
diagrammatic language, the complex numbers and other bosonic objects need no
strings attached. The rules of braided algebra are that sliding nodes under
strings etc (without cutting any strings) does not change the result of going
from the top to the bottom.

\begin{figure}
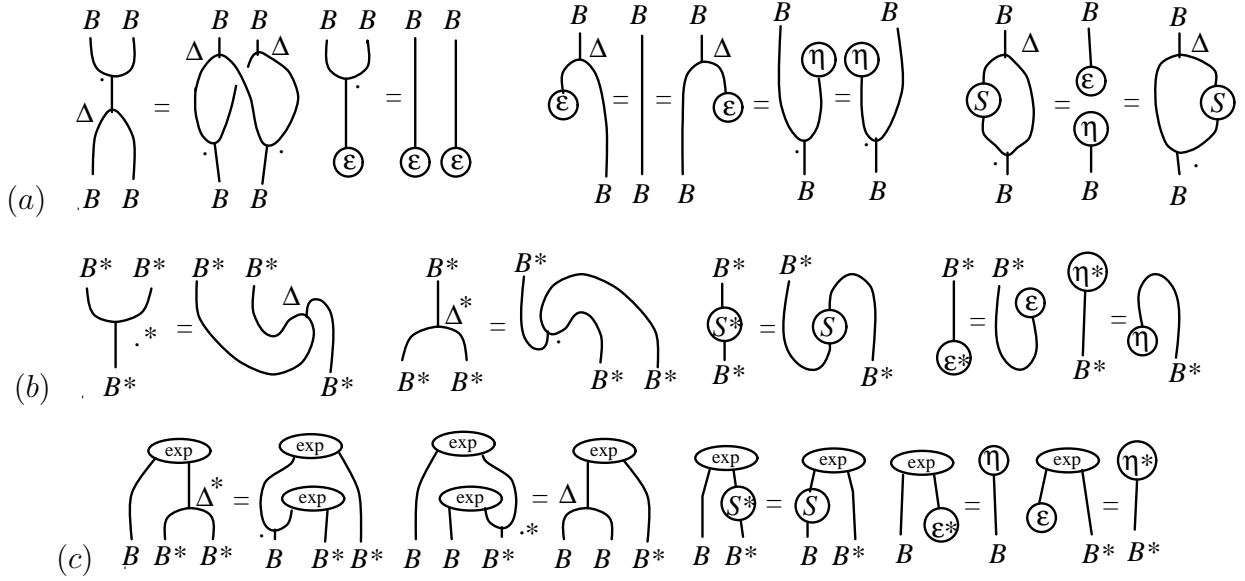

\[ (a)\quad \epsfbox{hopf-ax.eps}\]
\[ (b)\quad \epsfbox{B-dual.eps}\]
\[ (c)\quad \epsfbox{exp.eps}\]
\caption{(a) Axioms of a braided-Hopf algebra or `braided group' (b) Axioms for
dual braided group $B^*$ (c) The coevaluation map as an abstract exponential}
\end{figure}

We suppose that $B$ has a left dual Hopf algebra $B^*$ in the sense that there
is an evaluation pairing $\<\ ,\ \>=\epsfbox{cup.eps}:B^*\tens B\to \C$ obeying
certain properties. Namely, there should also be a {\em coevaluation}
$\coev=\epsfbox{cap.eps}:\C\to B\tens B^*$ such that we have the `double-bend
axioms' $\epsfbox{lbend.eps}=\vert$ as the identity map $B\to B$ and
$\epsfbox{rbend.eps}=\vert$ as the identity map $B^*\to B^*$. The product in
$B$ should be related to the coproduct in $B^*$ and vice-versa in the manner
recalled in Figure~1 (b). These elementary concepts from braided group theory
are all that we will need. An introduction to these concepts and methods is in
\cite{Ma:introp}.

Our first observation is an elementary one: applying the bend-straightening
axioms to the pairing of $B$ and $B^*$ in Figure~1(b), we obtain that
$\exp=\coev$ obeys
\ceqn{expcoprod}{ (\Delta\tens\id)\exp={\exp}_{23}{\exp}_{13},\quad
(\id\tens\Delta)\exp={\exp}_{13}{\exp}_{12}\\
(\eps\tens\id)\circ\exp=\eta,\quad (\id\tens\eps)\circ\exp=\eta}
which we have written in diagrammatic form in Figure~1(c). If we think of the
coproduct as `addition' in $B$ or $B^*$ (which will be exactly its role in our
examples) we see that the coevaluation always obeys the characteristic property
of an exponential.
If $\{e_a\}$ is any basis of $B$ and $\{f^a\}$ a dual basis then $\exp=\sum
e_a\tens f^a$ is the corresponding braided exponential. In the
infinite-dimensional case it means of course that $\exp$ is a formal
power-series, but one can still proceed by working order by order in a
deformation parameter etc in the manner well-known for the universal R-matrix
of a quantum group.

The role of the pairing $\<\ ,\ \>$ itself is to provide an action of $B^*$ on
$B$ by evaluation against the coproduct (the coregular representation) as
already explained in \cite{Ma:introp}. This action plays the role of
differentiation in our abstract picture. Thus the notion of duality of
braided-Hopf algebras has two pieces, evaluation and coevaluation. When we
think of $\Delta$ as `coaddition', these become differentiation and
exponentiation respectively. We will of course demonstrate all this concretely
in our examples in Sections~6.2 and~6.3.

Next, we assume that we have a left integral $\int:B\to \C$ on $B$. This is
required to obey
$(\id\tens\int)\circ\Delta=\eta\tens\int$, which is just the usual definition
of translation invariance under the coproduct. We also require that the map has
trivial braiding with other objects, i.e. can be represented as a map from $B$
into a free node. In practice, all our constructions are covariant under a
background quantum group which induces the braiding, and this last condition is
the assertion that the integral is invariant under the background quantum
group.

The left integral is the final ingredient we need for a Fourier transform. It
also allows one to define a new `convolution' product on $B$:

\begin{theorem} We introduce a {\em convolution product} on a braided-Hopf
algebra by
$*=(\int\circ\cdot \tens\id)\circ (S\tens\id)\circ\Delta$. Then $*$ is an
associative product $B\tens B\to B$.
\end{theorem}
\proof We write the definition of $*$ diagrammatically in the box in Figure~2
(a) and prove that it is associative. For the first equality we used the
coassociativity of $\Delta$, for the third that $S$ is a braided-anti-algebra
homomorphism, and for the fourth we used a lemma proven separately in part (b).
For the lemma we use the axioms in Figure~1 of a Hopf algebra to introduce an
antipode loop, and then the left-invariance of $\int$. \endproof

\begin{figure}
\[ \epsfbox{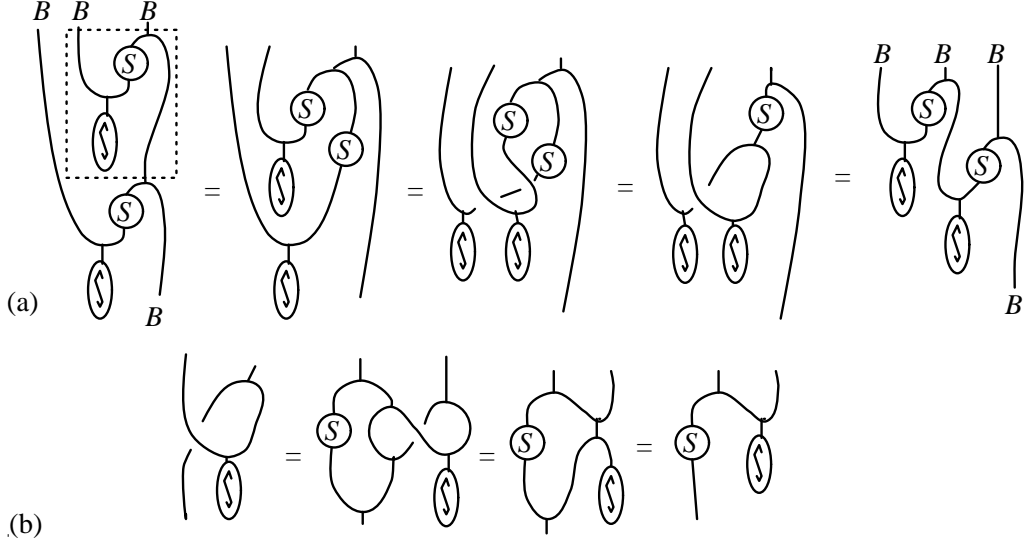}\]
\caption{(a) Proof of associativity of the convolution product (in box) for a
braided-Hopf algebra equipped with a left integra (b) Lemma needed in proof}
\end{figure}

Given a left integral, we define the abstract {\em Fourier transform operator}
$\fou:B^*\to B$ and abstract {\em delta-functions} on $B^*$ in the obvious way
by
\eqn{fou-delta}{\fou=(\int\circ\cdot\tens\id)\circ \exp,\quad
\delta^*=\fou\circ\eta=(\int\tens\id)\circ\exp.}

\begin{propos} The Fourier transform operator $\fou:B\to B^*$ intertwines the
standard right coregular representation of $B^*$ on $B$ in \cite{Ma:introp}
with the action by right multiplication in $B^*$.
\end{propos}
\proof This is given in Figure~3. The left upper box is the right coregular
action of $B^*$ on $B$ from \cite{Ma:introp}\cite{Ma:lie}. The lower left box
is $\fou$. We use the lemma in Figure~2(b). Then we use (\ref{expcoprod}) from
Figure~1(c). For the last equality we use $\exp=\epsfbox{cap.eps}$ and the
double-bend axiom for the coevaluation. \endproof

\begin{figure}
\[ \epsfbox{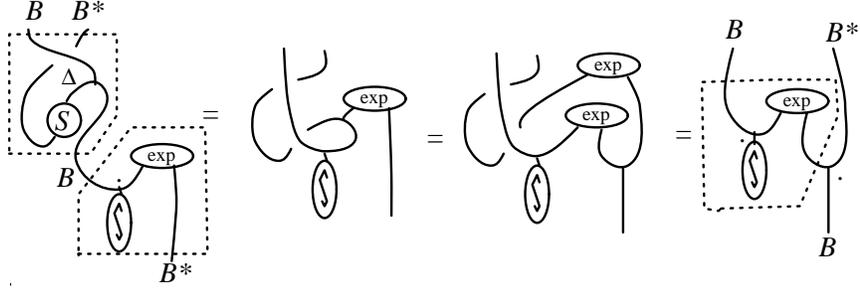}\]
\caption{Fourier transformation (lower box) maps the usual action of $B^*$ by
vector fields (upper box) to multiplication in $B^*$}
\end{figure}

The right-regular action here is the one introduced in \cite{Ma:introp} and
already computed for matrix braided groups in \cite{Ma:lie}. It corresponds to
the vector fields on $B$ generated by elements of $B^*$ acting by `translation'
on the underlying braided group, and always respects the product of $B$ (which
becomes a right braided-module algebra). The role of the antipode is to convert
the left action of these differential operators to an action from the right.
Proposition~6.2 is the fundamental property of Fourier transform.

Another useful property is
\eqn{deltafou}{ \Delta\circ\fou=
(\fou\tens\id)\circ(\cdot\tens\id)\circ(\id\tens\exp)}
which follows at once from (\ref{expcoprod}). This is useful for computing the
differential of a Fourier transform.
The Fourier transform also behaves well with respect to convolution:

\begin{theorem} The Fourier transform maps the opposite convolution product
from Theorem~6.1 for $B$ to the usual product of $B^*$.
\end{theorem}
\proof The opposite convolution product means $*\circ\Psi$ and is also
associative. The proof that $\fou$ is then an algebra homomorphism from this to
$B^*$ is in Figure~4. The second equality is (\ref{expcoprod}). The third is
the lemma in Figure~2(b). \endproof

\begin{figure}
\[ \epsfbox{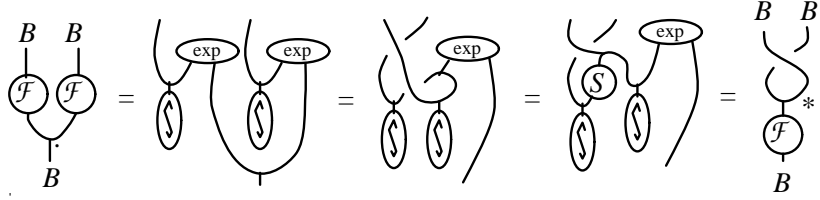}\]
\caption{Fourier transformation maps convolution $*\circ\Psi$ to the product of
$B^*$}
\end{figure}

To complete our picture we show that the Fourier transform is invertible. For
this we need to assume that we can also make a Fourier transform on $B^*$,
which we ensure by providing ourselves with a (right-handed) integral
$\int^*:B^*\to \C$ in addition to the integral above on $B$. If $B$ is our
quantum position space then $B^*$ is our quantum momentum space and we require
that we can integrate over this too. We define
\eqn{volfou*}{\vol=\int^*\delta^*=(\int\tens\int^*)\circ\exp,\quad
\fou^*=(\id\tens\int^*\circ\cdot)\circ(\Psi\tens\id)\circ(\id\tens\exp)}
where $\fou^*$ is the version of Fourier-transform on $B^*$ that we need. It is
a right-handed version (in contrast to our left-handed one above) converted to
a left-handed setting by $\Psi$.

\begin{propos}  $\fou^*\fou=S\, \vol$, where $S$ is the braided-antipode.
\end{propos}
\proof This is in Figure~5(a). The second equality is (\ref{expcoprod}), the
third is the lemma in Figure~2(b) and last is a useful lemma which we give
separately in part (b). In the  proof of the lemma we use the double-bend axiom
for $\exp=\epsfbox{cap.eps}$ and then (\ref{expcoprod}), followed by the
assumption that $\int^*$ is a right-integral. \endproof

\begin{figure}
\[ \epsfbox{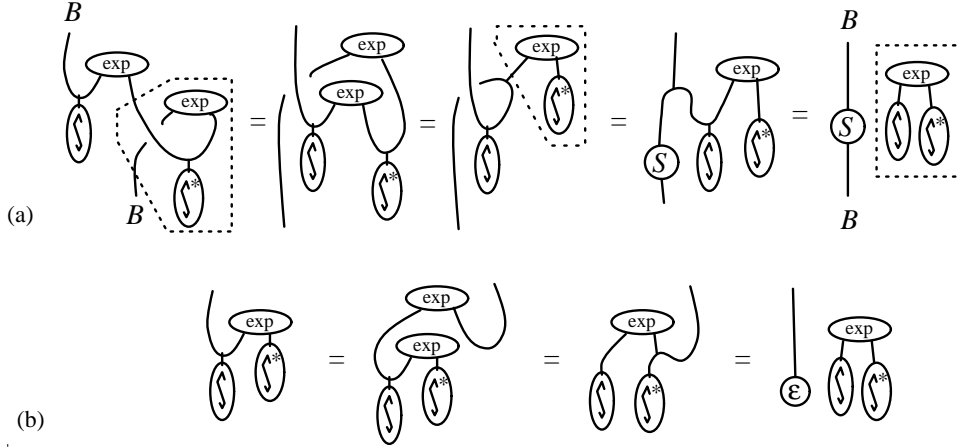}\]
\caption{(a) Fourier transform $\fou^*$ on $B^*$ (in left box) provides inverse
of $\fou$. Middle box is $\delta$-function on $B$ and right box is `volume'
element (b) Lemma needed in proof}
\end{figure}

One can show from this and a similar calculation that if $S$ is invertible then
so is $\fou$, with inverse $S^{-1}\circ \fou^*$. The braided-antipode $S$ here
plays the role of the minus sign in usual Fourier theory, while $\vol$ plays
the role of $2\pi$. One can also define a delta-function on $B$ by
$\delta=(\id\tens\int^*)\circ\exp$ and show that
\eqn{deltafun}{(\id\tens\int)(\Psi\tens\id)\circ\Delta\circ\delta=S\, \vol}
which is the characteristic evaluation property of delta-functions. This is
given in Figure~5(a) also as the right-hand half of the proof. One can show by
the same methods that if our integrals are both left and right invariant (the
nicest possible case) or at least are invariant under $S$, then $\delta$ is the
identity for the convolution product.

At this level of abstraction, our theory here works for Hopf algebras in any
braided category. This includes of course the (more standard) theory for
ordinary Hopf algebras, as well as \cite{LyuMa:bra}\cite{LyuMa:fou} for the
self-dual braided groups $B=\und H$ given by transmutation of quantum
enveloping algebras $H$. In the latter case
$\exp=(S\tens\id)(\CR_{21}\CR_{12})$ where $\CR$ is the universal R-matrix and
$S$ the ordinary antipode of $H$.

\subsection{1-dimensional case}

In this section we consider how these abstract constructions look in the
one-dimensional case where $B=\C[x]$ the functions in one formal variable $x$
as studied in Sections~1--4. This time we make full use of the Hopf algebra
structure as it was introduced in \cite{Ma:csta}, namely
\eqn{1coprod}{ \Delta x^m=\sum [{m\atop r};q] x^r\tens x^{m-r},\quad
Sx^m=(-1)^m q^{m(m-1)\over 2} x^m ,\quad \eps x^m=0.}
We use conventions with $q$ rather than $q^2$ as used before.

For the dually-paired Hopf algebra $B^*$ we take the same Hopf algebra with
variable $v$ say and analogous structure. Both live formally in the braided
category of $\Z$-graded algebras with $x$ of degree $|x|=1$ and $v$ of degree
$|v|=-1$. The braiding has a factor $q^{|x||v|}$ so that
\[ \Psi(x\tens x)=qx\tens x,\quad \Psi(v\tens v)=q v\tens v,\quad \Psi(x\tens
v)=q^{-1}v\tens x,\quad \Psi(v\tens x)=q^{-1}x\tens v.\]
This shows up when we consider two or more copies of the algebras. For example,
the braided tensor products $\C[x]\und\tens\C[y]$ with $y$ a copy of $x$,
$\C[v]\und\tens\C[w]$ with $w$ a copy of $v$ and $\C[x]\und\tens\C[v]$ have the
relations
\eqn{1btens}{ yx=qxy,\quad wv=qvw,\quad vx=q^{-1}xv}
respectively, since $vx\equiv(1\tens v)(x\tens 1)=\Psi(v\tens x)=q^{-1}x\tens
v\equiv q^{-1}xv$ etc. The commutation relations depend on exactly which
algebra one is working with (i.e. the order). For example,
$\C[v]\und\tens\C[x]$ is a different algebra. This is why the braiding notation
with $\Psi=\epsfbox{braid.eps}$ is useful as a way to keep track of the
$q$-factors in a completely coherent way. The coproduct above is by definition
the linear one extended as an algebra homomorphism to $\C[x]\und\tens \C[y]$,
so $\Delta f=f(x+y)$.

The pairing we take between $B,B^*$ is
\eqn{1pairing}{ \<f(v),g(x)\>=\eps\circ f(\del)g,\quad {\rm i.e.}\quad
\<v^m,x^n\>=\delta_{m,n}[m;q]!}
and the corresponding coevaluation, which is our abstract exponential, is
\eqn{1exp}{\exp=e_q^{x|v}=\sum_{m=0}^\infty {x^m v^m\over
[m;q]!}=\sum_{m=0}^\infty {(xv)^m\over [m;q^{-1}]!}\equiv e_{q^{-1}}^{xv}}
as an element of $\C[x]\tens\C[v]$. If we consider this with the braided tensor
product algebra
(\ref{1btens}) then $x(vx)=q(xv)x$ giving the right hand form if one wants to
work in this algebra. From the diagrammatic point of view, however, it is often
more convenient to keep the ordering with the symbol $|$ as on the left. One
has the properties for an exponential in Figure~1(c) as
\eqn{1expcop}{e_q^{x+y|v}=e_q^{y|v}e_q^{x(\ |\ )v}=e_q^{y|v}e_q^{x|v},\quad
e_q^{x|v+w}=e_q^{x(\ |\ )w}e_q^{x|v}=e_q^{x|w}e_q^{x|v}}
where the spaces are to insert the other factor in each term of the
exponential. For example, the first case lives in
$\C[x]\und\tens\C[y]\und\tens\C[v]$. Since $e_q^{y|v}$ is bosonic, it commutes
with $x$ in this algebra we can also write it as shown. Similarly for the
second half. Note also that it would be more conventional to consider $v$ as an
ordinary number, so that $\exp=e_q^{xv}$, but this would not work in the
general $n$-dimensional case and would also not be consistent with the second
half (\ref{1expcop}).

It is clear that $e_q^{x\del}f(y)=f(x+y)$, which implies in turn that
\eqn{1intinv}{ (\id\tens\int)\Delta f=\int^{y\infty}_{-y\infty} f(x+y)=1
\int^{y\infty}_{-y\infty} f(y),\quad \int^{y\infty}_{0} f(x+y)\equiv 1
\int^{y\infty}_{0} f(y)}
whenever our class of functions is considered to have the appropriate boundary
conditions. We will emphasize the first case but both are possible at our
algebraic level (and correspond to Fourier or Laplace transforms in appropriate
representations). This provides our left-invariant integral in position space
$x$. Note that we also required in the abstract theory that $\int$ be bosonic.
This is exactly the property of being bosonic in (\ref{scalinv}). For braiding
a second copy $y$ past $\intinf x$ gives $x$ a power of $q$, but the dependence
of the integral on the scaling parameter is exactly modulo such powers.

The abstract Fourier theory in Section~6.1 then looks as follows. The
convolution product, Fourier transform and the right coregular representation
$\ra$ are
\eqn{1conv}{ (f*g)(y)=\intinf x f(x) S_x g(x+y)}
\eqn{1fou}{ \fou(f)(v)=\intinf x f(x) e_q^{x|v}}
\eqn{1ra}{ f\ra g=\left(Sg(\del|L^{-1})\right)f=g(-\del\circ L^{-1})f}
where $S_x$ acts as in (\ref{1coprod}) on the $x$ variable and $L$ is the
scaling by $q$ operator.
The formula for $\ra$ here are from the upper box in Figure~3 as
\[ f \ra v^m= (-1)^m\del^m q^{m(m-1)\over 2}\circ L^{-m}(f)=(S\del^m) f\]
where the braiding of $f$ past $v^m$ is the origin of the $L^{-m}$ factor. We
then apply the antipode to $v^m$ and evaluate the pairing from (\ref{1pairing})
with $\Delta f$. The second form for $\ra$ uses $L^{-1}\circ\del=q\del\circ
L^{-1}$.

The fundamental property of Fourier transform from Proposition~6.2 comes out as
\eqn{fouinter}{ \fou(f\ra g)(v)=\fou(f)(v)g(v)}
One could verify this directly using the $q$-Leibniz rule and integration by
parts. Thus,
\[ \intinf x f(x) e_q^{x|v}v^m=\intinf x f(x) \del^m e_q^{x|v}
=\intinf x ((-\del\circ L^{-1})^mf)(x) e_q^{x|v}=\intinf x (f \ra v^m)
e_q^{x|v}\]

As a concrete demonstration of this braided Fourier calculus, we compute two
examples. For our first example we can start in the algebra
$\C[x]\und\tens\C[y]\und\tens\C[v]$ and
compute
\align{f(y)\ra e^{x|v}_q\equad &&=\sum^\infty_{m=0}{x^m\over [m,q]!} f(q^my)\ra
v^m= \sum_{m=0}^\infty {x^m\over [m,q]!}(S\del^m) f(y)\\
&&=\sum_{m=0}^\infty {(Sx^m)\over [m,q]!} \del^m f(y)= S_x e_q^{x|\del}f(y)
=S_x f(x+y)}
from the above. Alternatively, the same starting point can be viewed in
$\C[y]\tens \C[x]\tens \C[v]$ and computed from the upper left box in Figure~3
as follows. Since $e_q^{x|v}$ is bosonic, $f(y)$ braids past it without change.
We then apply $S_v$ to obtain $e_{q^{-1}}^{-x|v}\tens f(y)$. The coproduct and
evaluation via (\ref{1pairing}) gives $e_{q^{-1}}^{-x|\del}f(y)$ which is the
same as before. Either way, Proposition~6.2 then tells us that
\eqn{f-x1}{ \intinf y (S_x f(x+y)) e_q^{y|v}=\intinf y f(y) e_q^{y|v}
e_q^{x|v}.}
The lemma in Figure~2(c) is related to this and looks like
\eqn{f-x2}{ \intinf y (S_x f(x+y)) g(y)=\intinf y f(y)g(x+y)}
for general $f,g$. In these formulae, $S_x f(x+y)$ plays the role of $f(-x+y)$.

For a second example, we compute the Fourier transform of our Gau{\ss}ian
$g_\alpha$ from Section~4. Thus
\align{ \del_v\fou(g_\alpha)\equad&&=\del_v\intinf x g_\alpha e_q^{x|v}=\intinf
x g_\alpha x e_q^{x|v}\\
&&={1\over \alpha}\intinf x (-\del g_\alpha)  e_q^{x|v}={1\over
\alpha}(\fou\circ L(g_\alpha)) v =
{1\over \alpha}(L^{-1}\circ\fou (g_\alpha)) v}
so that $\fou(g_\alpha)$ is of the same general Gau{\ss}ian type (for a
slightly different operator) and with an inverted decay factor. One can obtain
the same result using derivatives acting from the right. In both cases the
action of the differential is immediate from (\ref{deltafou}), which looks as
\[ \fou(f)(v+w)=\fou(fe_q^{x|w}).\]
Here $w$ is understood to lie to the far right. This in turn is immediate from
(\ref{1expcop}).

To complete the picture, we assume now a similar but right-handed integral in
momentum space, i.e. such that
\eqn{1intinv*}{ (\int^*\tens\id)\Delta f\equiv \int_{-v\infty}^{v\infty}
f(v+w)=1 \int_{-v\infty}^{v\infty} f(v);\quad \int^{v\infty}_{0} f(v+w)=1
\int^{v\infty}_{0} f(v).}
Then
\eqn{1fou*}{ \fou^*(f)=\sum_{m=0}^\infty {x^m\over [m;q]!}\intinf v
L^{-m}(f)(v) v^m=\intinf v f(v) e_q^{x|v}}
\eqn{1fousquare}{\fou^*\fou(f)=\intinf v \intinf x f(x)  e_q^{x|v}e_q^{y|v}=
Sf(y) \intinf x \intinf v e_q^{x|v}=Sf(y) \, \vol}
where the first from of $\fou^*$ is the definition in Figure~5(a) while the
second is written assuming we are in the algebra $\C[x]\und\tens\C[v]$. The
first expression for $\fou^*\fou$ is its definition and Proposition~6.4
asserts that this coincides with the other expressions in our abstract setting.
Here $\intinf v e_q^{x|v}=\delta(x)$ is our abstract delta-function.

We have shown in this subsection that there is a coherent $q$-calculus for
Fourier transforms in one braided variable. One has to be careful about
ordering but this can be taken care of systematically by appealing to the
diagrammatic method.   One can further proceed to fix representations of the
system and examine convergence of limits and other issues normally associated
with harmonic analysis, for example with $x$ represented as a real number
and $v$ imaginary or vice versa.

\subsection{$n$-dimensional case}

We now give the general $n$-dimensional case of the above, i.e. in the same
level of generality as in Section~5.3. We take for $B$ the algebra of braided
covectors\cite{Ma:poi} generated by $\vecx=\{x_i\}$ as in (\ref{polys}) and for
$B^*$ the algebra of braided vectors generated by $\vecv=\{v^i\}$ with similar
relations $R'\vecv_2\vecv_1=\vecv_1\vecv_2$. We use the standard compact
notation (as in Theorem~5.1) where numerical suffices stand for the position of
the matrix or tensor indices.
Our goal is to see that the various conditions needed in Section~6.1 are
satisfied in this context. We use the differentials and braided-exponentials
from \cite{Ma:fre} and the integration theory from Section~5.3 above.

We need first the full braided-Hopf algebra structure of the braided covector
and vector algebras.
The required braid statistics are \cite{Ma:lin}
\eqn{vec-covec}{\vecy_1\vecx_2 =\vecx_2\vecy_1R,\quad \vecw_1\vecv_2
=R\vecv_2\vecw_1,\quad \vecv_1\vecx_2=\vecx_2 R^{-1}\vecv_1,\quad \vecx_1
R\vecv_2=\vecv_2\vecx_1}
if $\vecy$ is a second copy of the covector algebra and $\vecw$ a second copy
of the vector algebra. These relations correspond to the braiding
\ceqn{nbraiding}{\Psi(\vecx_1\tens \vecx_2)=\vecx_2\tens \vecx_1R,\quad
\Psi(\vecv_1\tens \vecv_2)=R\vecv_2\tens \vecv_1\nonumber\\
\Psi(\vecv_1\tens\vecx_2)=\vecx_2 \tens R^{-1}\vecv_1,\quad \Psi(\vecx_1 \tens
R\vecv_2)=\vecv_2\tens \vecx_1}
Recall that in the framework of \cite{Ma:lin}, everything is manifestly
covariant under $\vecx\to \vecx\vect \dila$ and $\vecv\to
\dila^{-1}\vect^{-1}\vecv$ when there is an associated quantum group with
matrix generator $\vect$ and dilaton $\dila$. It plays the role of the
$\Z$-grading in the 1-dimensional case and induces the braiding $\Psi$ between
any two comodule-algebras in a coherent way. The dilaton is needed when the
quantum group normalization factor $\lambda$ is not $1$.

Keeping in mind such braiding $\Psi$ or the corresponding braid statistics
relations, one can add vectors etc, along familiar lines. Thus
$\vecx+\vecy$ obeys the correct relations and statistics with other objects, if
$\vecx,\vecy$ do\cite{Ma:poi}\cite{Ma:fre}. Abstractly, there is a braided-Hopf
algebra structure
\eqn{n-coprod}{\Delta \vecx=\vecx\tens1+1\tens\vecx,\quad
S\vecx=-\vecx,\quad\eps\vecx=0}
extended via the above braiding. The higher powers are given by braided
R-binomial coefficients\cite{Ma:fre}. Similarly for $\vecv+\vecw$. These are
the main ideas of braided-linear algebra as introduced by the second author. We
refer to \cite{Ma:introp} for an introduction.

Next, we recall the basic formulae for R-differentiation and R-exponentiation
introduced in \cite{Ma:fre}. Namely, following the same line of reasoning as in
the braided derivation of the $q$-derivative, one finds
\eqn{deli}{\del^i(\vecx_1\cdots \vecx_m)= {\bf e}^i{}_1\vecx_2\cdots\vecx_m
\left[m;R\right]_{1\cdots m}}
where ${\bf e}^i$ is a basis covector $({\bf e}^i){}_j=\delta^i{}_j$ and
\[\left[m;R\right]=1+(PR)_{12}+(PR)_{12}(PR)_{23}
+\cdots+(PR)_{12}\cdots (PR)_{m-1,m}\]
is a certain {\em braided integer} matrix\cite{Ma:fre} living in the $m$-fold
matrix tensor product. Here $P$ is the usual permutation matrix. The
one-dimensional case $R=(q)$ recovers the usual $q$-derivative, while the
2-dimensional case for the $SL_2$ R-matrix recovers \cite{WesZum:cov}. It was
also shown in \cite{Ma:fre} that the $\del^i$ obey the relations of a
braided-vector algebra (like the $v^i$) and the braided Leibniz rule
(\ref{Rleibniz}).

Crucial for us now is that this differentiation defines a duality pairing of
braided vectors and covectors by
\eqn{pairing}{ \<f(\vecv),g(\vecx)\>=\eps\circ f(\del)g(\vecx).}
With this pairing, we can say that the braided vectors $\vecv$ are the braided
Hopf-algebra dual of the braided covectors\cite{Ma:introp}. We assume that the
pairing is non-degenerate, which is true for standard deformations and other
generic R-matrices. In this case, we are in the abstract setting of
Section~6.1. The coevaluation likewise exists in the general case, as a formal
power-series.

Also introduced in \cite{Ma:fre} was a formula for the R-exponential or
coevaluation in this setting. This was given explicitly in the universal or
free case where $R'=P$ so that no relations at all need be assumed among the
$x_i$ or among the $v^i$ separately. Then\cite{Ma:fre}
\eqn{exp}{\exp_R(\vecx|\vecv)=\sum_{m=0}^{\infty}
\vecx_1\cdots\vecx_m\left[m;R\right]_{1\cdots
m}^{-1}\cdots \left[2;R\right]_{m-1,
m}^{-1}\vecv_m\cdots\vecv_1}
is an eigenfunction of $\del^i$ and generates finite translations\cite{Ma:fre}
\eqn{eigen-Taylor}{\del^i\exp_R(\vecx|\vecv)=\exp_R(\vecx|\vecv)v^i,\quad
\exp_R{(\vecx|\del)}f(\vecy)=f(\vecx+\vecy).}
Here the eigenvalues $v^i$ are elements of a non-commutative algebra, such as
the vector algebra where the $\del^i$ live.
Finally, one has also the characteristic properties \cite{Ma:fre}
\eqn{expaddn}{ \exp_R(\vecx+\vecy|\vecv)=\exp_R(\vecy|\vecv)\exp_R(\vecx(\ |\
)\vecv),\quad \exp_R(\vecx|\vecv+\vecw)=\exp_R(\vecx(\ |\
)\vecw)\exp_R(\vecx|\vecv)}
where  $(\ |\ )$ denotes a space for $\exp_R(\vecy|\vecv)$ and
$\exp_R(\vecx|\vecw)$ respectively to be inserted in each term of the
exponential. In fact, the R-exponential is invariant under the transformation
by $\vect$ and hence bosonic in the sense that  its braid-statistics with
anything else is trivial. So we can also write (\ref{expaddn}) more simply as
in (\ref{expcoprod}) and (\ref{1expcop}) if the appropriate braided tensor
product algebra is understood.
Finally, $\exp$ fits together with the pairing (\ref{pairing}) as
\eqn{rigid}{\<f(\vecv),\exp_R(\vecx|\vecw)\>= f(\vecw),\quad
\<\exp_R(\vecx|\vecv),g(\vecy)\>=g(\vecx)}
since it is an eigenfunction of the $\del^i$ operators. Here the pairing in the
second case is between $\vecv$ and $\vecy$.

It is obvious that in the one-dimensional case with $R=(q)$, the universal
R-exponential collapses to the more usual $q$-exponential in the previous
subsection. Moreover, the general braided spaces with $R'$ are quotients of the
$n$-dimensional free one, and the appropriate $\exp_R$ is obtained by
projecting down our universal one. The only subtlety here is that
$F(m;R)=[m;R]^{-1}\cdots [2;R]^{-1}$ need not be invertible as soon as there
are relations among the $x_i$: all that is really needed in the proofs in
\cite{Ma:fre} is that
\[ \vecx_2\cdots\vecx_m[m;R]_{1\cdots
m}F(m;R)\vecv_m\cdots\vecv_1=\vecx_2\cdots\vecx_mF(m-1;R)\vecv_m\cdots\vecv_1\]
For example, an R-matrix is called {\em Hecke} if
\eqn{hecke}{ R_{21}=q^2 R^{-1}+(q^2-1)P}
and in this case one has\cite{Ma:poi} a braided vector space with $R'=q^{-2}R$.
Since $PR'\vecv_2\vecv_1=\vecv_2\vecv_1$, we know without any calculation that
$F(m;R)=F(m;q^{2})=[m;q^2]!$ in this case. Hence the $q$-exponential introduced
in \cite{Ma:fre} collapses in the Hecke case to
\[ \exp_{R}(\vecx|\vecv)=\sum_{m=0}^\infty
{\vecx_1\cdots\vecx_m\vecv_m\cdots\vecv_1\over
[m;q^2]!}=e_{q^{-2}}^{\vecx\cdot\vecv}.\]
Here the second form follows trivially from $\vecx_1
(\vecx\cdot\vecv)=q^2(\vecx\cdot\vecv)\vecx_1$, which is valid in the Hecke
case as an easy consequence of the braid-statistics relations
(\ref{vec-covec}):
\[\vecx_1 (\vecx_2\cdot\vecv_2)=q^{-2}\vecx_2\vecx_2
R\vecv_2=q^{-2}\vecx_2\vecx_1(q^2
R_{21}^{-1}+(q^2-1)P)\vecv_2=(\vecx_2\cdot\vecv_2)\vecv_1+(1-q^{-2})
\vecx_1(\vecx_2\cdot\vecv_2).\]
This second Hecke form of the R-exponential (\ref{exp}) was stressed recently
in \cite{ChrZum:tra}, who also repeated some of the general derivation of
\cite{Ma:fre} in terms of $\vecx\cdot\vecv$. We would like to stress that this
collapsing of the R-exponential into an ordinary $q$-exponential is only valid
in the simplest cases such as the Hecke one. Some physically interesting
braided vector spaces, such as $q$-Minkowski space do have an addition
law\cite{Mey:new}, but the relevant $R$ is not Hecke. In this important case,
U. Meyer has observed that  $\exp_R(\vecx|\vecv)=e_{q^2}^{\vecx|\vecv}$
nevertheless holds if $\vecv$ is a null vector\cite{Mey:wav}.

Next, we need a left-invariant integration. Here we use $\int$ as constructed
implicitly in Theorem~5.1 for these braided covector spaces. We gave the
construction for general $R,R'$ and an invariant metric $\eta_{ij}$ subjects to
some reasonable conditions. Invariance of the metric ensures that $\int$ is
invariant too. This means that it is bosonic in  the sense that it has trivial
braiding, as we assumed in Section~6.1. To know $\int$ explicitly, we need to
assume that the corresponding Gau{\ss}ian and its inverse can be constructed.
Assuming this, we know by construction that the integral of $\del^i f$ vanish.
Hence from (\ref{eigen-Taylor}) we conclude global translation invariance too.
Alternatively, one can formulate all our theory directly in terms of the
invariant linear functional $\Zint$ and consider expressions with $\int$ only
as a useful notational. We can now conclude the basic Fourier theory in this
setting too. We have
\eqn{n-conv}{f*g(\vecy)=\int f(\vecx)S_{\vecx} g(\vecx+\vecy)}
\eqn{n-fou}{\fou(f)(\vecv)=\int f(\vecx)\exp_R(\vecx|\vecv)}
with $\fou$ an algebra homomorphism from $*\circ\Psi$ to the pointwise product
for $\vecv$. We also have the fundamental property
\eqn{n-fund}{\fou(f\ra g)(\vecv)=\fou(f)(\vecv) g(\vecv)}
where $\ra$ is the right coregular action of $\vecv$. There are also similar
constructions for a right-integral on the braided vector algebras which then
provides for the inverse Fourier transform.

It remains only to compute the right coregular action $\ra$ in our setting. We
do this from the upper left box in Figure~3. From (\ref{nbraiding}) we have
\eqn{xrav}{ x_i\ra v^j=\tilde{R}^a{}_i{}^j{}_b
\<(-v^b),x_a\>=-\tilde{R}^a{}_i{}^j{}_a\equiv -u^j{}_i,\quad{\rm i.e.}\quad
\vecx_1 R\ra \vecv_2=-\id}
where the first form is explicit and the second is in the compact notation. We
note that the matrix $u$ here implements the square of the antipode of the
background quantum group. For the more general case we compute
\ceqn{psixxv}{\Psi(\vecx_1\cdots\vecx_m\tens v^i)=v^a\tens\vecx_1\cdots\vecx_m
\left(\tilde{R}_{m\ m+1}\cdots\tilde{R}_{1\ m+1}\right)^i{}_a\\
\Psi(\vecx_1\cdots\vecx_m R_{1\ m+1}\cdots R_{m\ m+1}\tens
\vecv_{m+1})=\vecv_{m+1}\tens\vecx_1\cdots\vecx_m.}
The braiding here is obtained in the manner explained in \cite{Ma:lin}. The
pairing (\ref{pairing}) then gives $(\vecx_1\cdots\vecx_m)\ra v^i$ in terms of
$\tilde{R}$ or
\eqn{xxxrav}{(\vecx_1\cdots\vecx_m R_{1\ m+1}\cdots R_{m\ m+1})\ra
\vecv_{m+1}=-\del_{m+1}(\vecx_1\cdots\vecx_m)}
in terms of $R$. We know from the theory of braided groups in \cite[Sec.
4]{Ma:introp} that $\ra$ is necessarily a right action of the braided vector
algebra and extends to products of the $\vecx$ as a braided module-algebra. In
the present case this means that it is a right-handed braided-derivation
\eqn{vleib}{ \left(f(\vecx)g(\vecx)\right)\ra v^i=f(\vecx)\ra\Psi(g(\vecx)\tens
v^i)+f(\vecx)(g(\vecx)\ra v^i).}
One can also take a Heisenberg algebra point of view as we did in Section~5 for
the usual derivatives. Following the construction analogous to \cite{Ma:fre}
for the usual $\del^i$, we have
\eqn{vheis}{ \lvec {x_i} \lvec {v^j}-\tilde{R}^b{}_i{}^j{}_a \lvec {v^a} \lvec
{x_b}=-u^j{}_i,\quad {\rm i.e.}\quad \lvec {v^i}\lvec {x_j}  - x_a
R^a{}_j{}^i{}_b \lvec {v^b}=\delta^i{}_j}
as a right-handed version of (\ref{Rleibniz}), where $\lvec {x_i},\lvec {v^j}$
denote operators acting on the position co-ordinates from the right, by
multiplication and $\ra$ respectively.

Some applications of this theory will be explored elsewhere. In particular, we
see that we now have the main ingredients needed to do  classical field theory
and elements of quantum field theory (such as computing braided-Feynman
diagrams) on  braided spaces, as part of a general programme of $q$-deforming
physics.

\section{Appendix}
Let us examine the Gau{\ss}ian function in the quasiposition
representations of Section~3. It is the solution of
\be
(\pax + \alpha {\bf x})\v{\ga} = 0
\ee
where
\be
\v{\ga} = \sum_{s=-\infty}^{+\infty} {\ga}_s \v{v_{cq^{2s}}}.
\ee
{}From
\be
\pax + \alpha {\bf x} = \sum_r \left[ \v{v_{cq^{2r}}}
\left( \frac{1}{cq^{2r}(1-q^2)} + \alpha c q^{2r} \right)
\langle v_{cq^{2r}} \vert - \v{v_{cq^{2r}}}
\frac{1}{cq^{2r}(1-q^2)}\langle v_{cq^{2(r+1)}} \vert \right]
\ee
we get
\be
{\ga}_{s+1} = (1+ \alpha c q^{4s} (1-q^2) ) {\ga}_s
\ee
The normalisation of $\ga$ be $\langle v_c\v{\ga} = 1$. Thus
\be
{\ga}_r = \prod_{n=1}^r (1 + \alpha c q^{4n} (1-q^2))
\mbox{ \qquad for } r=1,2,...
\ee
which is convergent for $r\rightarrow \infty$.
Similarly one obtains
\be
{\ga}_{r} = \prod_{n=1}^{-r} \frac{1}{1 + \alpha c q^{-4n} (1-q^2)}
\mbox{ \qquad for } r= -1,-2,...
\ee
which is convergent to zero for $r\rightarrow -\infty$.
One can easily check that we even have
\be
(x^t \ga)_r = c q^{2tr}
\prod_{n=1}^{-r} \frac{1}{1 + \alpha c q^{-4n} (1-q^2)}
\quad \rightarrow 0 \mbox{ \qquad for } r \rightarrow -\infty
\ee
which proves the rapid decay property of $\ga$.

Let us now also check its integrability:
\align{&&\equad \int_{-c\infty}^{c\infty} \ga = 2 (1-q^2) \mbox{Trace}({\bf
x}\ga)\\
&&= 2 c (1-q^2)\left[1/2+\sum_{r=1}^{\infty} q^{2r}
\prod_{n=1}^r (1 + \alpha c q^{4n} (1-q^2))
+ \sum_{r=1}^{\infty} q^{-2r}
\prod_{n=1}^{r} (1 + \alpha c q^{-4n} (1-q^2))^{-1}
\right]}
Both the infrared and the ultraviolet part of the integral
are easily checked to be convergent by using a ratio test.


\baselineskip 20pt

\begin{thebibliography}{10}
\itemsep 0pt

\bibitem{Ma:exa}
S.~Majid.
\newblock Examples of braided groups and braided matrices.
\newblock {\em J. Math. Phys.}, 32:3246--3253, 1991.

\bibitem{Ma:lin}
S.~Majid.
\newblock Quantum and braided linear algebra.
\newblock {\em J. Math. Phys.}, 34:1176--1196, 1993.

\bibitem{Ma:poi}
S.~Majid.
\newblock Braided momentum in the {$q$}-{P}oincar{\'e} group.
\newblock {\em J. Math. Phys.}, 34:2045--2058, 1993.

\bibitem{Ma:fre}
S.~Majid.
\newblock Free braided differential calculus, braided binomial theorem and the
  braided exponential map.
\newblock {\em J. Math. Phys.}, 34:4843--4856, 1993.

\bibitem{Ma:lie}
S.~Majid.
\newblock Quantum and braided {L}ie algebras, {F}ebruary 1993 (damtp/93-4).
\newblock {\em J. Geom. Phys.}
\newblock To appear.

\bibitem{Mey:new}
U.~Meyer.
\newblock A new {$q$}-{L}orentz group and $q$-{M}inkowski space with both
  braided coaddition and {$q$}-spinor decomposition.
\newblock {\em Preprint}, DAMTP/93-45, 1993.

\bibitem{Ma:euc}
S.~Majid.
\newblock {$q$}-{E}uclidean space and quantum {W}ick rotation by twisting,
  {J}anuary 1993 (damtp/94-03).
\newblock {\em J. Math. Phys.}, January, 1994.
\newblock To appear.

\bibitem{Ma:introp}
S.~Majid.
\newblock Beyond supersymmetry and quantum symmetry (an introduction to braided
  groups and braided matrices).
\newblock In M-L. Ge and H.J. de~Vega, editors, {\em Quantum Groups, Integrable
  Statistical Models and Knot Theory}, pages 231--282. World Sci., 1993.

\bibitem{Exton}
H.~Exton.
\newblock {\em {$q$}-{H}ypergeometric Functions and Application}.
\newblock Ellis Horwood Ltd., Chichester, U.K., 1983.

\bibitem{And:ser}
G.E. Andrews.
\newblock {$q$}-{S}eries: Their development and application in analysis, number
  theory, combinatorics, physics and computer algebra.
\newblock Technical Report~66, AMS, 1986.

\bibitem{Ma:csta}
S.~Majid.
\newblock {$\C$}-statistical quantum groups and {W}eyl algebras.
\newblock {\em J. Math. Phys.}, 33:3431--3444, 1992.

\bibitem{Jac:int}
F.H. Jackson.
\newblock {$q$}-{I}ntegration.
\newblock {\em Proc. Durham Phil. Soc.}, 7:182--189, 1927.

\bibitem{BauFlo:pat}
L.~Baulieu and E.G. Floratos.
\newblock Path integral on the quantum plane.
\newblock {\em Phys. Lett. B}, 258:171--178, 1991.

\bibitem{Kem:sym}
A.~Kempf.
\newblock Quantum group-symmetric {F}ock-spaces and {B}argmann-{F}ock
  representation.
\newblock {\em Lett. Math. Phys.}, 26:1--12, 1992.

\bibitem{KorVak:spa}
L.I. Korogodsky and L.L. Vaksman.
\newblock Quantum {$G$}-spaces and {H}eisenberg algebra.
\newblock In {\em Proc. of the Euler Institute, St. Petersberg (1990)}, volume
  1510 of {\em Lec. Notes. in Math.}, pages 56--66. Springer.

\bibitem{ChrZum:tra}
C.~Chryssomalakos and B.~Zumino.
\newblock Translations, integrals and fourier transforms in the quantum plane.
\newblock {\em Preprint}, November, 1993.

\bibitem{HebWei:fre}
A.~Hebecker and W.~Weich.
\newblock Free particle in $q$- deformed configuration space.
\newblock {\em Lett. Math. Phys.}, 26:245--258, 1992.

\bibitem{Fio:sym}
G.~Fiore.
\newblock The {$SO_q(N,R)$}-symmetric harmonic oscillator on the quantum
  {E}uclidean space {$R^N_q$} and its {H}ilbert space structure.
\newblock {\em Int. J. Mod. Phys. A}, 26:4679--4729, 1993.

\bibitem{WesZum:cov}
J.~Wess and B.~Zumino.
\newblock Covariant differential calculus on the quantum hyperplane.
\newblock {\em Proc. Supl. Nucl. Phys. B}, 18B:302, 1990.

\bibitem{Kem:int}
A.~Kempf.
\newblock Quantum group-symmetric {F}ock-spaces and {B}argmann-{F}ock space:
  Integral kernels, green functions, driving forces.
\newblock {\em J. Math. Phys.}, 34:969--987, 1993.

\bibitem{FRT:lie}
L.D. Faddeev, N.Yu. Reshetikhin, and L.A. Takhtajan.
\newblock Quantization of {L}ie groups and {L}ie algebras.
\newblock {\em Leningrad Math J.}, 1:193--225, 1990.

\bibitem{Dri}
V.G. Drinfeld.
\newblock Quantum groups.
\newblock In A.~Gleason, editor, {\em Proceedings of the {ICM}}, pages
  798--820, Rhode Island, 1987. AMS.

\bibitem{Ma:add}
S.~Majid.
\newblock On the addition of quantum matrices.
\newblock {\em J. Math. Phys.}, 35:2617--2633, 1994.

\bibitem{CWSSW:lor}
U.~Carow-Watamura, M.~Schlieker, M.~Scholl, and S.~Watamura.
\newblock A quantum {L}orentz group.
\newblock {\em Int. J. Mod. Phys.}, 6:3081--3108, 1991.

\bibitem{OSWZ:def}
O.~Ogievetsky, W.B. Schmidke, J.~Wess, and B.~Zumino.
\newblock {$q$}-{D}eformed {P}oincar{\'e} algebra.
\newblock {\em Comm. Math. Phys.}, 150:495--518, 1992.

\bibitem{MaMar:glu}
S.~Majid and M.~Markl.
\newblock Glueing operation for {$R$}-matrices, quantum groups and link
  invariants of {H}ecke type.
\newblock {\em Preprint}, DAMTP/93-20, 1993.

\bibitem{LyuMa:bra}
V.V. Lyubashenko and S.~Majid.
\newblock Braided groups and quantum {F}ourier transform.
\newblock {\em J. Algebra}, 166(3), 1994.

\bibitem{LyuMa:fou}
V.V. Lyubashenko and S.~Majid.
\newblock Fourier transform identities in quantum mechanics and the quantum
  line.
\newblock {\em Phys. Lett. B}, 284:66--70, 1992.

\bibitem{Mey:wav}
U.~Meyer.
\newblock Wave equations on $q$-{M}inkowski space.
\newblock {\em Preprint}, DAMTP/94-10, 1994.

\end{thebibliography}
\end{document}